\documentclass[nojss,article]{jss}
\usepackage[]{graphicx}
\usepackage[]{color}
\makeatletter
\def\maxwidth{ %
  \ifdim\Gin@nat@width>\linewidth
    \linewidth
  \else
    \Gin@nat@width
  \fi
}
\makeatother

\usepackage{Sweave}

\usepackage{amsmath}
\usepackage{natbib}
\usepackage[latin1]{inputenc}
\usepackage[T1]{fontenc}
\usepackage[english]{babel}

\author{Jouni Helske\\ University of Jyv\"askyl\"a}
\title{\pkg{KFAS}: Exponential Family State Space Models in \proglang{R}}

\Plainauthor{Jouni Helske} 
\Plaintitle{KFAS: Exponential Family State Space Models in R} 
\Shorttitle{\pkg{KFAS}: Exponential Family State Space Models in \proglang{R}} 

\Abstract{
State space modelling is an efficient and flexible method for statistical inference of a broad class of time series and other data. This paper describes an \proglang{R} package \pkg{KFAS} for state space modelling with the observations from an exponential family, namely Gaussian, Poisson, binomial, negative binomial and gamma distributions. After introducing the basic theory behind Gaussian and non-Gaussian state space models, an illustrative example of Poisson time series forecasting is provided. Finally, a comparison to alternative \proglang{R} packages suitable for non-Gaussian time series modelling is presented.
}

\Plainkeywords{R, exponential family, state space models, time series, forecasting, dynamic linear models} 

\Address{
Jouni Helske\\
University of Jyv\"askyl\"a\\
Department of Mathematics and Statistics\\
40014 Jyv\"askyl\"a, Finland\\
E-mail: \email{Jouni.Helske@jyu.fi}
}

\begin{document}

\section{Introduction}

State space models offer a unified framework for modelling several types of time series and other data. Structural time series, autoregressive integrated moving average (ARIMA) models, simple regression, generalized linear mixed models, and cubic spline smoothing are just some examples of the statistical models which can be represented as a state space model. One of the simplest classes of state space models are linear Gaussian state space models (also known as dynamic linear models), which are analytically tractable, and are therefore often used in many fields of science.

\citet{Petris2010} and \citet{Tusell2010} introduce and review some of the contributed \proglang{R} \citep{R} packages available at Comprehensive \proglang{R} Archive Network (CRAN) for Gaussian state space modelling. Since then, several new additions have emerged in CRAN. Most of these packages use one package or multiple packages reviewed in \citet{Tusell2010} for filtering and smoothing and add new user interfaces and functionality for certain type of models. For example, package \pkg{rucm} \citep{rucm} is focused on structural time series, \pkg{dlmodeler} \citep{dlmodeler} provides a unified interface compatible with multiple packages, and \pkg{MARSS} \citep{MARSSarticle,MARSS} provides functions for the  maximum likelihood estimation of a large class of Gaussian state space models via the EM-algorithm.

One of the packages reviewed in the aforementioned papers is \pkg{KFAS} (the Kalman Filtering And Smoothing). Besides of modelling the general linear Gaussian state space models, \pkg{KFAS} can also be used in cases where the observations are from other exponential family models, namely binomial, Poisson, negative binomial, and Gamma models. 

After the papers by \citet{Petris2010} and \citet{Tusell2010}, \pkg{KFAS} has been completely rewritten. The package is now much more user-friendly due to the use of \proglang{R}'s symbolic formulas in model definition. The non-Gaussian modelling, which was somewhat experimental in the old versions of \pkg{KFAS}, is now fully functional supporting multivariate models with different distributions. Many other features have also been added (such as methods for computing model residuals and predictions), the performance of the main functions has improved and in the process several bugs have been fixed.

In this paper I first introduce the basic theory related to state space modelling, and then proceed to show the main aspects of \pkg{KFAS} in more detail, illustrate its functionality by applying it to real life dataset, and finally make a short comparison between \pkg{KFAS} and other potentially useful packages for non-Gaussian time series modelling.

\section{Gaussian state space model}

In this section an introduction to key concepts regarding the theory of Gaussian state space modelling as in \pkg{KFAS} is given. As the algorithms behind \pkg{KFAS} are mostly based on \citet{DK2012} and the related articles by the same authors, the basic notation is nearly identical with the one used by Durbin and Koopman. 

For the linear Gaussian state space model with continuous states and discrete time intervals $t=1,\ldots,n,$ we have
\begin{equation}\label{ssgeneral}
\begin{aligned}
y_t &= Z_t\alpha_t + \epsilon_t, \quad \textrm{(observation equation)}\\
\alpha_{t+1} &= T_t\alpha_t + R_t\eta_{t}, \quad \textrm{(state equation)}
\end{aligned}
\end{equation}
where $\epsilon_{t} \sim N(0,H_t), \eta_{t} \sim N(0,Q_t)$ and $\alpha_1 \sim
N(a_1,P_1)$ independently of each other. We assume that $y_t$ is a $p\times 1$,
$\alpha_{t+1}$ is an $m \times 1$ and $\eta_{t}$ is a $k \times 1$ vector. We also denote $\alpha = (\alpha^\top_1,\ldots,\alpha^\top_n)^\top$ and similarly $y = (y^\top_1,\ldots,y^\top_n)^\top$.

Here $y_t$ contains the observations at time $t$, whereas $\alpha_t$ is a vector of latent state process at time point $t$. The system matrices $Z_t$, $T_t$, and $R_t$, together with the covariance matrices $H_t$ and $Q_t$ depend on the particular model definition, and are often time invariant, i.e., do not depend on $t$. Usually at least some of these matrices contain unknown parameters which need to be estimated. In \pkg{KFAS} one defines the model with the function \code{SSModel}. The function \code{SSModel} only builds the model and does not perform estimation of unknown parameters, which differs from functions like \code{lm}, which builds and estimates the model with one command.

The main goal of the state space modelling is to gain knowledge of the latent states $\alpha$ given the observations $y$. This is achieved by using two important recursive algorithms, the Kalman filtering and smoothing. From the Kalman filtering algorithm we obtain the one-step-ahead predictions and the prediction errors
\begin{equation*}
\begin{aligned}
a_{t+1} &= \E(\alpha_{t+1}|y_t,\ldots,y_1), \\
v_{t} &= y_t - Z_t a_t
\end{aligned}
\end{equation*}
and the related covariance matrices
\begin{equation*}
\begin{aligned}
P_{t+1} &= \VAR(\alpha_{t+1}|y_t,\ldots,y_1), \\
F_{t} &= \VAR(v_t) = Z_t P_t Z_t^\top + H_t.
\end{aligned}
\end{equation*}

Using the results of the Kalman filtering, we establish the state smoothing equations running backwards in time and yielding
\begin{equation*}
\begin{aligned}
\hat\alpha_{t} &= \E(\alpha_{t}|y_n,\ldots,y_1),  \\
V_{t} &= \VAR(\alpha_{t}|y_n,\ldots,y_1).
\end{aligned}
\end{equation*}
Similar smoothed estimates can also be computed for the disturbance terms $\epsilon_t$ and $\eta_t$, and straightforwardly for the signal $\theta_t=Z_t\alpha_t$. For details on these algorithms, see Appendix~\ref{appendix} and \citet{DK2012}.

A prior distribution of the initial state vector $\alpha_1$ can be defined as a multivariate Gaussian distribution with mean $a_1$ and covariance matrix $P_1$. For an uninformative diffuse prior, one typically sets $P_1 = \kappa\textrm{I}$, where $\kappa$ is $10^7$, for example. However, this method can be numerically unstable due to cumulative roundoff errors. To solve this issue \citet{KD2003} present the exact diffuse initialization method, where the diffuse elements in $a_1$ are set to zero and $P_1$ is decomposed as $\kappa P_{\infty,1} + P_{*,1}$, where $\kappa\to\infty$. Here $P_{\infty,1}$ is a diagonal matrix with ones on those diagonal elements which relate to the diffuse elements of $\alpha_1$, and $P_{*,1}$ contains the covariances of the nondiffuse elements of $\alpha_1$ (and zeros elsewhere). At the start of the Kalman filtering (and at the end of backward smoothing) we use so-called exact diffuse initialisation formulas until $P_{\infty,t}$ becomes a zero matrix, and then continue with the usual Kalman filtering equations. This exact method should be less prone to numerical errors, although they can still occur especially in the smoothing phase, if we, for example, have high collinearity between the explanatory variables of the model. Note that given all the parameters in the system matrices, results from the Kalman filter and smoother are equivalent with Bayesian analysis given the same prior distribution for $\alpha_1$.

When we have multivariate observations, it is possible that in the diffuse phase the matrix $F_t$ is not invertible, and the computation of $a_{t+1}$ and $P_{t+1}$ becomes impossible. On the other hand, even if $F_t$ is invertible, the computations can become slow when the dimensionality of $F_t$, that is, the number of series increases. Also in the case of multivariate observations, the formulas relating to the diffuse initialization become cumbersome. Based on the ideas of \citet{AM1979}, a complete univariate approach for filtering and smoothing was introduced by \citet{KD2000} (known as sequential processing by Anderson and Moore). The univariate approach is based on the alternative representation of the model~\eqref{ssgeneral}, namely
\begin{equation*}
\begin{aligned}
y_{t,i} &= Z_{t,i}\alpha_{t,i} + \epsilon_{t,i},  \quad i=1,\ldots,p_t, \quad t=1,\ldots,n, \\
\alpha_{t,i+1} &= \alpha_{t,i},  \quad i=1,\ldots,p_t-1,\\
\alpha_{t+1,1} &= T_t\alpha_{t,p_t} + R_t\eta_{t},  \quad t=1,\ldots,n,
\end{aligned}
\end{equation*}
and $a_{1,1} \sim N(a_1,P_1)$, with the assumption that $H_t$ is diagonal for all $t$. Here the dimension of the observation vector $y_t$ can vary over time and therefore missing observations are handled straightforwardly by adjusting the dimensionality of $y_t$. In the case of non-diagonal $H_t$, the original model can be transformed either by taking the LDL decomposition of $H_t$, and multiplying the observation equation with the $L_t^{-1}$, so $\epsilon_t^* \sim N(0,D_t)$, or by augmenting the state vector with $\epsilon$, when $Q_t$ becomes block diagonal with blocks $Q_t$ and $H_t$. Augmenting can also be used for introducing a correlation between $\epsilon$ and $\eta$. Both the LDL decomposition and the state vector augmentation are supported in \pkg{KFAS}.

In theory, when using the univariate approach, the computational costs of filtering and smoothing
decrease, as the number of matrix multiplications decrease, and there is no need for solving the system of equations \citep[p. 159]{DK2012}. As noted in \citet{Tusell2010}, these gains can somewhat cancel out as more calls to linear algebra functions are needed and the memory management might not be as effective as working with larger objects at once. Nevertheless, as noted previously, sequential processing has also other clear benefits, especially with diffuse initialization where the univariate approach simplifies the recursions considerably \citep{DK2012}.

\pkg{KFAS} uses this univariate approach in all cases. Although $v_t$, $F_t$, and $K_t = P_t Z_t^\top = \COV(a_t,y_t|y_{t-1},\ldots,y_1)$ differ from the standard multivariate versions, we get $a_t=a_{t,1}$ and $P_t=P_{t,1}$ by using the univariate approach. If standard multivariate matrices $F_t$ and $K_t$ are needed for inference, they can be computed later from the results of the univariate filter. As the covariances $F_{\ast,i,t}$, $K_{\ast,i,t}$, and $P_{\ast,t}$ relating to the diffuse phase (see Appendix~\ref{appendix}) coincide with the
nondiffuse counterparts if $F_{\infty,i,t}=0$, the asterisk is dropped from the variable names in \pkg{KFAS}, and, for example, the variable \code{F} is an $n \times p$ array containing $F_{\ast,i,t}$ and $F_{i,t}$, whereas \code{Finf} is an $n \times d$, where $d$ is the last time point before the diffuse phase ended.

\subsection{Log-likelihood of the Gaussian state space model}

The Kalman filter equations can be used for computing the log-likelihood, which
in its standard form is
\begin{equation*}
\log L  = -\frac{np}{2}\log2\pi - \frac{1}{2}\sum^n_{t=1}(\log|F_t| +
v^\prime_tF^{-1}_tv_t).
\end{equation*}
In the case of the univariate treatment and diffuse initialization, the diffuse
log-likelihood can be written as
\begin{equation*}
\begin{aligned}
\log L_d  &= -\frac{1}{2}\sum^{n}_{t=1}\sum^{p_t}_{i=1}w_{i,t},
\end{aligned}
\end{equation*}
where
\begin{displaymath}
w_{i,t} = \left\{ \begin{array}{ll}
\log F_{\infty,i,t}, &\mbox{if $F_{\infty,i,t}>0$,} \\
I(F_{i,t}>0)(\log2\pi + \log F_{i,t} + v^2_{i,t}F_{i,t}^{-1}), &\mbox{if $F_{\infty,i,t}=0$}.
\end{array} \right.
\end{displaymath}
See \citet[Chapter 7]{DK2012} for details. \citet{Francke2010} show that there are cases where the above definition of diffuse log-likelihood is not optimal. Without going into the details, if system matrices $Z_t$ or $T_t$ contain unknown parameters in their diffuse parts, the diffuse likelihood is missing one term which depends on those unknown parameters. \citet[p.411--412]{Francke2010} present a recursive formula for computing this extra term, which is also supported by \pkg{KFAS}.

\subsection{Example of Gaussian state space model}\label{gaussian_example}

Now the theory of previous sections is illustrated via example. Our time series consists of yearly alcohol-related deaths per 100,000 persons in Finland for the years 1969--2007 in the age group of 40--49 years (Figure~\ref{fig:gaussian_plot}). The data is taken from \citet{stat1, stat2}.

For the observations $y_1, \ldots, y_n$ we assume that $y_t \sim N(\mu_t, \sigma_\epsilon)$ for all $t=1,\ldots,n$, where $\mu_t$ is a random walk with drift process
\begin{equation*}
\begin{aligned}
\mu_{t+1} &=  \mu_t + \nu + \eta_t\\
\end{aligned}
\end{equation*}
with $\eta_{t} \sim N(0,\sigma^2_{\eta})$. Assume that we have no prior information about the initial state $\mu_1$ or the constant slope $\nu$. This model can be written in a state space form by defining
\begin{displaymath}
Z =\left( \begin{array}{cc}
1 & 0
\end{array}\right),
H = \sigma_{\epsilon}^2,
T =
\left( \begin{array}{cc}
1 & 1 \\
0 &1  
\end{array} \right),
\end{displaymath}
\begin{displaymath}
\alpha_t =
\left( \begin{array}{c}
\mu_t \\
\nu_t
\end{array}\right),
R =
\left( \begin{array}{c}
1 \\
0 \\
\end{array}\right),
Q=\sigma^2_{\eta},
\end{displaymath}
\begin{displaymath}
a_1 =
\left( \begin{array}{c}
0 \\
0
\end{array}\right),
P_{*,1} =
\left( \begin{array}{cc}
0 & 0 \\
0 & 0
\end{array} \right),
P_{\infty,1} =
\left( \begin{array}{cc}
1 & 0 \\
0 & 1
\end{array} \right).
\end{displaymath}

In \pkg{KFAS}, this model can be written with the following code. For illustrative purposes we define all the system matrices manually without resorting default values.

\begin{Schunk}
\begin{Sinput}
R> data("alcohol")
R> deaths <- window(alcohol[, 2], end = 2007)
R> population <- window(alcohol[, 6], end = 2007)
R> Zt <- matrix(c(1, 0), 1, 2)
R> Ht <- matrix(NA)
R> Tt <- matrix(c(1, 0, 1, 1), 2, 2)
R> Rt <- matrix(c(1, 0), 2, 1)
R> Qt <- matrix(NA)
R> a1 <- matrix(c(1, 0), 2, 1)
R> P1 <- matrix(0, 2, 2)
R> P1inf <- diag(2)
R> 
R> model_gaussian <- SSModel(deaths / population ~ -1 + 
+      SSMcustom(Z = Zt, T = Tt, R = Rt, Q = Qt, a1 = a1, P1 = P1, 
+        P1inf = P1inf), 
+    H = Ht)
\end{Sinput}
\end{Schunk}
The first argument to the \code{SSModel} function is the formula which defines the observations (left side of tilde operator \code{~}) and the structure of the state equation (right side of tilde). Here \code{deaths / population} is a univariate time series, and the state equation is defined using the system matrices with auxiliary function \code{SSMcustom}, and the intercept term is omitted with \code{-1} in order to keep the model identifiable. The observation level variance is defined via the argument \code{H}. The \code{NA} values represent the unknown variance parameters $\sigma_{\epsilon}^2$ and $\sigma_{\eta}^2$ which can be estimated using the function \code{fitSSM}. After estimation, the filtering and smoothing recursions are performed using the \code{KFS} function.
\begin{Schunk}
\begin{Sinput}
R> fit_gaussian <- fitSSM(model_gaussian, inits = c(0, 0), method = "BFGS")
R> out_gaussian <- KFS(fit_gaussian$model)
\end{Sinput}
\end{Schunk}
In this case, the maximum likelihood estimates are 9.5 for $\sigma_{\epsilon}^2 $ and 4.3 for $\sigma_{\eta}^2$.

From the Kalman filter algorithm we get one-step-ahead predictions for the states $a_t = (\mu_t, \nu_t)^{\top}$. Note that even though the slope term $\nu$ was defined as time-invariant ($\nu_t = \nu$) in our model, it is recursively estimated by the Kalman filter. Thus at each time point $t$ when the new observation $y_t$ becomes available, the estimate of $\nu$ is updated to take account of the new information given by $y_t$. At the end of Kalman filtering, $a_{n+1}$ gives our final estimate of the constant slope term given all of our data. Here the slope term is estimated as 0.84 with standard error 0.34. For $\mu_t$, the Kalman filter gives the one-step-ahead predictions, but as the state is time-varying, we need to run also the smoothing algorithm if we are interested in the estimates of $\mu_t$ for $t=1,\ldots,n$ given all the data. 

Figure~\ref{fig:gaussian_plot} shows the observations with one-step-ahead predictions (red) and smoothed (blue) estimates of the random walk process $\mu_t$. Notice the typical pattern; at the time $t$ the Kalman filter computes the one-step-ahead prediction error $v_t = y_t - \mu_t$, and uses this and the the previous prediction to correct the prediction for the next time point (see Appendix~\ref{appendix} for the detailed update formula). Here this is most easily seen at the beginning of the series where our predictions seem to be lagging the observations by one time step. On the other hand, the smoothing algorithm takes account of both the past and the future values at each time point, thus producing more smoothed estimates of the latent process.

\begin{Schunk}
\begin{figure}[!ht]

{\centering \includegraphics[width=0.7\linewidth]{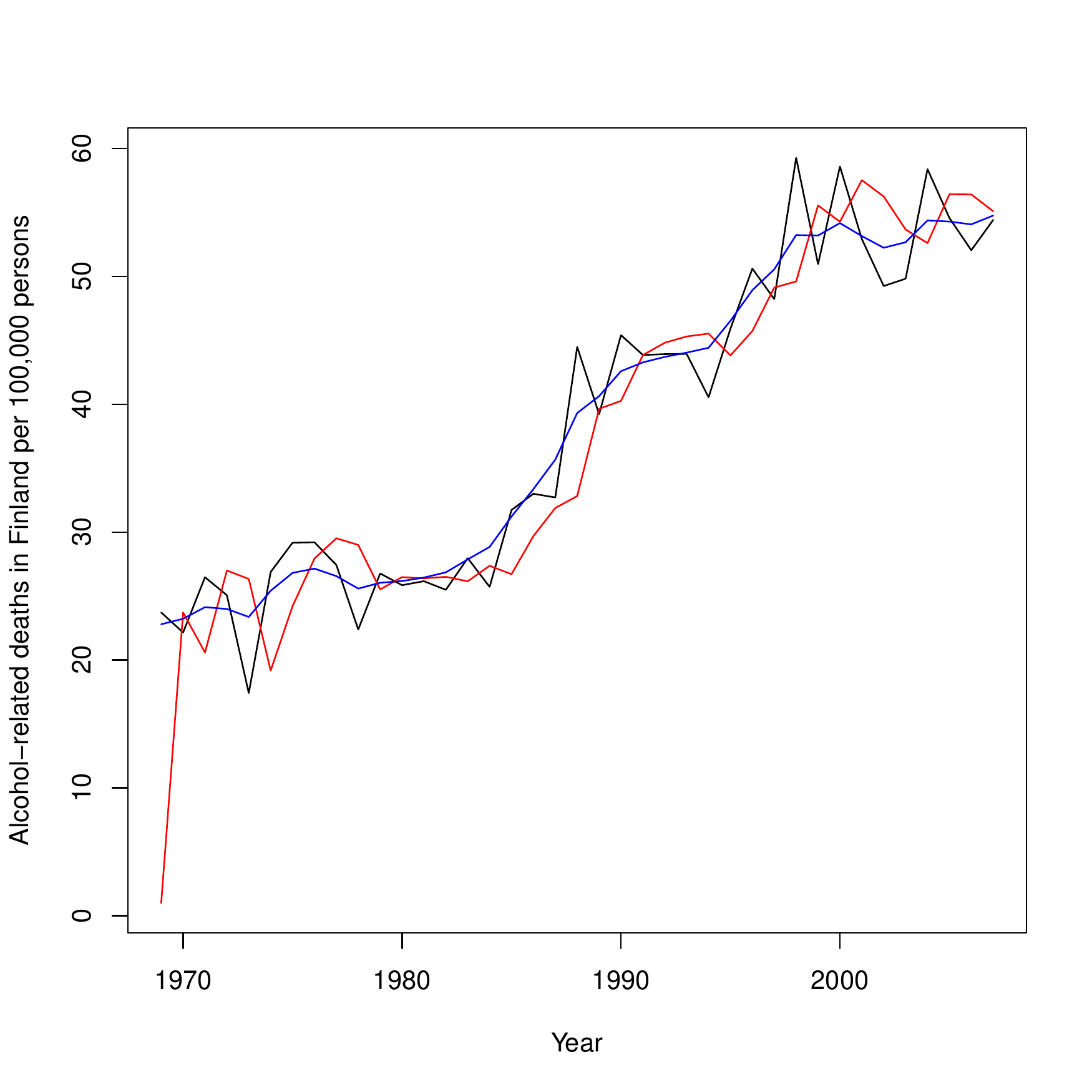} 

}

\caption[Alcohol-related deaths in Finland in the age group of 40--49 years (black line) with predicted (red) and smoothed (blue) estimates]{Alcohol-related deaths in Finland in the age group of 40--49 years (black line) with predicted (red) and smoothed (blue) estimates.}\label{fig:gaussian_plot}
\end{figure}
\end{Schunk}

\section{State space models for the exponential family}\label{exp}

\pkg{KFAS} can also deal with observations which come from distributions of an exponential family class other than Gaussian. We assume that the state equation is as in the Gaussian case, but the observation equation has the form
\begin{equation*}
p(y_t|\theta_t) = p(y_t|Z_t\alpha_t),
\end{equation*}
where $\theta_t=Z_t\alpha_t$ is the signal and $p(y_t|\theta_t)$ is the observational density.

The signal $\theta_t$ is the linear predictor which is connected to the expected value $E(y_t)=\mu_t$ via a link function $l(\mu_t)=\theta_t$. In \pkg{KFAS}, the following distributions and links are available:

\begin{enumerate}
\item Gaussian distribution with mean $\mu_t$ and variance $u_t$ with identity link $\theta_t=\mu_t$.

\item Poisson distribution with intensity $\lambda_t$ and exposure $u_t$ together with log-link $\theta_t = \log(\lambda_t)$. Thus we have $\E(y_t|\theta_t)=\VAR(y_t|\theta_t)=u_t e^{\theta_t}$.

\item Binomial distribution with size $u_t$ and probability of success $\pi_t$. \pkg{KFAS} uses logit-link so $\theta_t = \textrm{logit}(\pi_t)$ resulting $\E(y_t|\theta_t)=u_t\pi_t$ and $\VAR(y_t|\theta_t) = u_t(\pi_t(1-\pi_t))$.

\item Gamma distribution with a shape parameter $u_t$ and an expected value $\mu_t$, again with log-link $\theta_t = \log(\mu)$, where Gamma distribution is defined as
\begin{equation*}
p(y_t|\mu_t,u_t) = \frac{u_t^{u_t}}{\Gamma(u_t)}\mu_t^{-u_t}y_t^{u_t-1}e^{\frac{y_t u_t}{\mu_t}}.
\end{equation*}
This gives us $\E(y_t|\theta_t)=e^{\theta_t}$ and $\VAR(y_t|\theta_t) = e^{2\theta_t}/u_t$.

\item Negative binomial distribution with a dispersion parameter $u_t$ and an expected value $\mu_t$ with log-link $\theta_t = \log(\mu_t)$, where the negative binomial distribution is defined as
\begin{equation*}
p(y_t|\mu_t,u_t) = \frac{\Gamma(y_t+u_t)}{\Gamma(u_t)y_t!}\frac{\mu_t^{y_t} u_t^{u_t}}{(\mu_t+u_t)^{u_t+y_t}},
\end{equation*}
giving us $\E(y_t|\theta_t)=e^{\theta_t}$ and $\VAR(y_t|\theta_t) = e^{\theta_t} + e^{2\theta_t}/u_t$.
\end{enumerate}

Note that the variable $u_t$ has a different meaning depending on the distribution it is linked to. In \pkg{KFAS} one defines the distribution for each time series via argument \code{distribution} and the additional known parameters $u_t$ corresponding to each series as columns of the matrix \code{u}.

In order to make inferences of the non-Gaussian models, we first find a Gaussian model which has the same conditional posterior mode as $p(\theta|y)$ \citep{DK2000}. This is done using an iterative process with Laplace approximation of $p(\theta|y)$, where the updated estimates for $\theta_t$ are computed via the Kalman filtering and smoothing from the approximating Gaussian model. In the approximating Gaussian model the observation equation is replaced by
\begin{equation*}
\tilde y_t = Z_t\alpha_t + \epsilon_t, \quad \epsilon_t \sim N(0,H_t)
\end{equation*}
where the pseudo-observations $\tilde y_t$ variances $H_t$ are based on the first and second derivatives of $\log p(y_t|\theta_t)$ with respect to $\theta_t$ \citep{DK2000}.

Final estimates $\hat \theta_t$ correspond to the mode of $p(\theta|y)$. In the Gaussian case the mode is also the mean. In cases listed in (1)-(5) the difference between the mode and the mean is often negligible. Nevertheless, we are usually more interested in $\mu_t$ than in the linear predictor $\theta_t$. As the link function is non-linear, direct transformation $\hat \mu_t=l^{-1}(\hat \theta_t)$ introduces some bias. To solve this problem \pkg{KFAS} also contains methods based on importance sampling, which allows us to correct these possible approximation errors. With the importance sampling technique we can also compute the log-likelihood and the smoothed estimates for $f(\alpha)$, where $f$ is an arbitrary function of states, $\exp(Z_t\alpha_t)$ being a typical example.

In the importance sampling scheme, we first find the approximating Gaussian model, simulate the states $\alpha^i$ from this Gaussian model and then compute the corresponding weights $w_i=p(y|\alpha^i)/g(y|\alpha^i)$, where $p(y|\alpha^i)$ represents the conditional non-Gaussian density of the original observations, and $g(y|\alpha^i)$ is the conditional Gaussian density of the pseudo-observations $\tilde y$. These weights are then used for computing
\begin{equation*}
\E(f(\alpha)|y) = \frac{\sum_{i=1}^N f(\alpha^i) w_i}{\sum_{i=1}^Nw_i}.
\end{equation*}

The simulation of Gaussian state space models in \pkg{KFAS} is based on the simulation smoothing algorithm by \citet{DK2002}. In order to improve simulation efficiency, \pkg{KFAS} can use two antithetic variables in the simulation algorithms. See \citet[p.~265-266]{DK2012} for details on how these are constructed.

\pkg{KFAS} also provides means for the filtering of non-Gaussian models. This is achieved by sequentially using the smoothing scheme for $(y_1,\ldots,y_t), t=1\ldots,n$ with $y_t$ set as missing. This is a relatively slow procedure for large models, as the importance sampling algorithms need to be performed $n$ times, although the first steps are much faster than the one using the whole data. The non-Gaussian filtering is mainly for the computation of recursive residuals (see Section~\ref{residuals}) and for illustrative purposes, where computational efficiency is not that important. With large models or online-filtering problems, one is recommended to use a proper particle filter approach, which is out of the scope of this paper.

For non-Gaussian exponential family models in the context of generalized linear models, a typical way of obtaining the \textit{confidence} interval of the prediction is to compute confidence intervals in the scale of a linear predictor, and then the interval is transformed to the scale of observations. The issue of prediction intervals is often dismissed. For obtaining proper prediction intervals in the case of non-Gaussian state space models, the following algorithm is used in \pkg{KFAS}.
\begin{enumerate}
\item[(1)] Draw $N$ replicates of the linear predictor $\theta$ from the approximating Gaussian density $g(\theta|y)$ with importance weights $p(y | \theta)/g(y | \theta)$. Denote this sample $\tilde \theta^1,\ldots,\tilde \theta^N$ as $\tilde \theta$
\item[(2)] Using the importance weights as sampling probabilities, draw a sample of size $N$ with replacement from $\tilde \theta$. We now have $N$ independent draws from $p(\theta|y)$.
\item[(3)] For each $\tilde \theta^i$ sampled in step (2), take a random sample of $y^i$ from the observational distribution $p(y|\theta^i)$.
\item[(4)] Compute the the prediction intervals as empirical quantiles from $y^1,\ldots,y^N$.
\end{enumerate}

Assuming all the model parameters are known, these intervals coincide (within the Monte Carlo error) with the ones obtained from Bayesian analysis using the same priors for states.

\subsection{Log-likelihood of the non-Gaussian state space model}

The log-likelihood function for the non-Gaussian model can be written as \citep[p.~272]{DK2012}
\begin{equation*}
\begin{aligned}
\log L(y) &= \log \int p(\alpha,y)\textrm{d}\alpha \\
&= \log L_g(y)+ \log E_g\left[\frac{p(y|\theta)}{g(y|\theta)}\right],
\end{aligned}
\end{equation*}
where $L_g(y)$ is the log-likelihood of the Gaussian approximating model and the expectation is taken with respect to the Gaussian density $g(\alpha|y)$. The expectation can be approximated by
\begin{equation}\label{eg}
\log E_g\left[\frac{p(y|\theta)}{g(y|\theta)}\right] \approx \log\frac{1}{N}\sum_{i=1}^N w_i.
\end{equation}

In many cases, a good approximation of the log-likelihood can be computed without any simulation, by setting $N=0$ and using the mode estimate $\hat \theta$ from the approximating model.

In practice~\eqref{eg} suffers from the fact that $w_i= p(y|\theta^i)/g(y|\theta^i)$ is numerically unstable; when the number of observations is large, the discrete probability mass function $p(y|\theta^i)$ tends to zero, even when the Gaussian density function $g(y|\alpha^i)$ does not. Therefore it is better to redefine the weights as
\begin{equation*}
w^*_i = \frac{p(y|\theta^i)/p(y|\hat\theta)}{g(y|\theta^i)/g(y|\hat \theta)}.
\end{equation*}

The log-likelihood is then computed as
\begin{equation*}
\begin{aligned}
\log \hat L(y) &= \log L_g(y) + \log \hat w + \log\frac{1}{N}\sum_{i=1}^Nw^*_i,
\end{aligned}
\end{equation*}
where $\hat w = p(y|\hat \theta)/g(y|\hat \theta)$.

\subsection{Example of non-Gaussian state space model}\label{nongaussian_example}

The alcohol-related deaths of Section~\ref{gaussian_example} can also be modelled naturally as a Poisson process. Now our observations $y_t$ are the actual counts of alcohol-related deaths in year $t$, whereas the varying population size is taken account of by the exposure term $u_t$. The state equation remains the same, but the observation equation is now of form $p(y_t|\mu_t) = \textrm{Poisson}(u_t e^{\mu_t})$. 

\begin{Schunk}
\begin{Sinput}
R> model_poisson <- SSModel(deaths ~ -1 + 
+      SSMcustom(Z = Zt, T = Tt, R = Rt, Q = Qt, P1inf = P1inf), 
+    distribution = "poisson", u = population)
\end{Sinput}
\end{Schunk}
Compared to the Gaussian model of Section~\ref{gaussian_example}, we now need to define the distribution of the observations using the argument \code{distribution} (which defaults to \code{"gaussian"}). We also define the exposure term via the argument \code{u} (for non-Gaussian models the \code{H} is omitted and vice versa), and use default values for \code{a1} and \code{P1} in the \code{SSMcustom}.

In this model there is only one unknown parameter, $\sigma^2_{\eta}$. This is estimated as 0.0053, but the actual values of $\sigma^2_{\eta}$ between the Gaussian and Poisson models are not directly comparable as the intepretation of $\mu_t$ differs between models. The slope term of the Poisson model is estimated as 0.022 with standard error \ensuremath{1.4\times 10^{-4}}, corresponding to the 2.3\% yearly increase in deaths.

Figure~\ref{fig:nongaussian_example} shows the smoothed estimates of the intensity (deaths per 100,000 persons) modelled as Gaussian process (blue), and as a Poisson process (red).

\begin{Schunk}
\begin{figure}[!ht]

{\centering \includegraphics[width=0.7\linewidth]{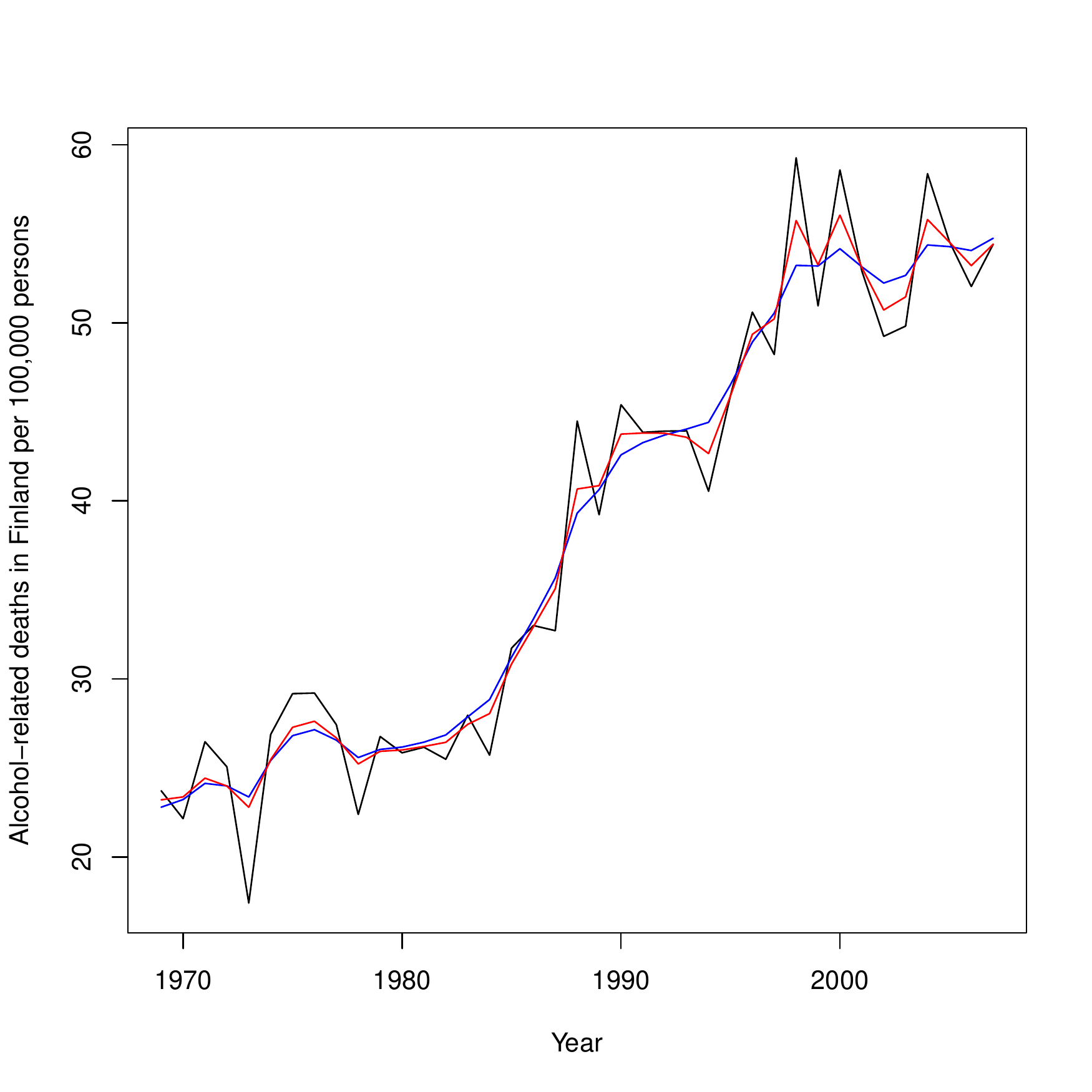} 

}

\caption[Alcohol-related deaths in Finland (black line) with smoothed estimates from Gaussian model (blue) and Poisson model (red)]{Alcohol-related deaths in Finland (black line) with smoothed estimates from Gaussian model (blue) and Poisson model (red).}\label{fig:nongaussian_example}
\end{figure}
\end{Schunk}

\section{Residuals}\label{residuals}

For exponential family state space models, multiple types of residuals can be computed. Probably the most useful ones are standardized recursive residuals, which are based on the one-step-ahead predictions from the Kalman filter. For the univariate case these are defined as
\begin{equation*}\label{recres}
\frac{y_t-\E(y_t|y_{t-1},\ldots,y_1)}{\sqrt{\VAR(y_t|y_{t-1},\ldots,y_1)}},\quad t = d+1\ldots,n,
\end{equation*}
where $d$ is the last time point of the diffuse phase, and the denominator can be decomposed as
\begin{equation*}
\begin{aligned}
\VAR(y_t|y_{t-1},\ldots,y_1)&= \VAR(\E(y_t|\theta_t,y_{t-1},\ldots,y_1)|y_{t-1},\ldots,y_1)\\
&+ \E(\VAR(y_t|\theta_t,y_{t-1},\ldots,y_1)|y_{t-1},\ldots,y_1)\\
&= \VAR(\E(y_t|\theta_t)| y_{t-1},\ldots,y_1) + \E(\VAR(y_t|\theta_t)|y_{t-1},\ldots,y_1) .
\end{aligned}
\end{equation*}
In the Gaussian case this simplifies to $v_{t}F_{t}^{-\frac{1}{2}}$.

For multivariate observations we have several options on how to standardize the residuals. The most common one is a marginal standardization approach, where each residual series is divided by its standard deviation, so we get residual series which should not exhibit any autocorrelations. Another option is to use, for example, Cholesky decomposition for the prediction error covariance matrix $F_t$ and standardize the residuals by $L_t^{-1}(y_t-\hat y_t)$ where $L_t L_t^\top=F_t$. Now the whole series of residuals (treated as a single univariate series) should not contain any autocorrelation.

For computing the marginally standardized residuals, multivariate versions of $F_t$ and $v_t$ are needed, whereas the Cholesky standardized residuals can be computed directly from the sequential Kalman filter as
\begin{equation*}
v_{i,t} F_{i_t}^{-\frac{1}{2}}, \quad j=1,\ldots,p,\quad t=d+1\ldots,n.
\end{equation*}
These multivariate residuals depend on the ordering of the series, so if the residual diagnostics exhibit deviations from model assumptions, then the interpretation is somewhat more difficult than when using the marginal residuals. Therefore marginal residuals might be preferred. Note that if we want quadratic form residuals $(y_t- \hat y_t)F_t^{-1}(y_t- \hat y_t)$, then the ordering of the series does not matter.

The recursive residuals are defined just for the non-diffuse phase, which is problematic if the model contains a long diffuse phase, for example, because a dummy variable with a diffuse prior is incorporated to the model. This is because the diffuse phase cannot end before the dummy variable changes its value at least once. In order to circumvent this, one can use a proper but highly non-informative prior distribution for the intervention variable when computing the residuals, which should have a negligible effect on the visual inspection of the residual plots.

Other potentially useful residuals are auxiliary residuals, which are based on smoothed values of states. For details, see \citet{HarveyKoopman1992} and \citet[Chapter 7]{DK2012}.

\section[Functionality of KFAS]{Functionality of \pkg{KFAS}}

The state space model used with \pkg{KFAS} is built using the function \code{SSModel}. The function uses \proglang{R}'s formula object in a similar way to that of the functions \code{lm} and \code{glm}, for example. In order to define the different components of the state space model, auxiliary functions \code{SSMtrend}, \code{SSMseasonal}, \code{SSMcycle}, \code{SSMarima}, \code{SSMregression} are provided. These functions can be used to define the structural, ARIMA, and regression components of the model. The function \code{SSMcustom} can be used for constructing an arbitrary component by directly defining the system matrices of the model~\eqref{ssgeneral}. More details on how to construct common state space models with \pkg{KFAS} are presented in Section~\ref{models}.

The function \code{SSModel} returns an object of class \code{SSModel}, which contains the observations \code{y} as the \code{ts} object, system matrices \code{Z},\code{H},\code{T},\code{R},\code{Q} as arrays of appropriate dimensions, together with matrices \code{a1}, \code{P1}, and \code{P1inf} defining the initial state distribution. Additional components contain the system matrix \code{u} which is used in non-Gaussian models for additional parameters, the character vector \code{distribution} which defines the distributions of the observations (multivariate series can have different distributions), and the tolerance parameter \code{tol} which is used in diffuse phase for checking whether $F_\infty$ is nonzero.

\code{SSModel} object also contains some attributes, namely, integer valued attributes \code{p},\code{m},\code{k}, and \code{n} which define the dimensions of the system matrices, character vectors \code{state_types} and \code{eta_types} which define the elements of $\alpha_t$ and $\eta_t$, and integer vector \code{tv} which defines whether the model contains time-varying system matrices. These attributes are used internally by \pkg{KFAS}, although the user can carefully modify them if needed. For example, if the user wishes to redefine the error term $\eta_t$ by changing the dimensions of \code{R} and \code{Q}, the attributes \code{k} and \code{eta_types} need to be updated accordingly.

The unknown model parameters can be estimated with \code{fitSSM}, which is a wrapper around \pkg{R}'s \code{optim} function and the \code{logLik} method for the \pkg{SSModel} object. For \code{fitSSM}, the user gives the model object, initial values of unknown parameters and a function \code{updatefn}, which is used to update the model given the parameters (the help page of \code{fitSSM} gives an example of \code{updatefn}). As the numerical optimization routines update the model and compute the likelihood thousands of times, the user is encouraged to build his own problem-specific model updating function for maximum efficiency. By default, \code{fitSSM} estimates the \code{NA} values in the time invariant covariance matrices $H$ and $Q$, but no general estimation function is provided. Of course, the user can also directly use the \code{logLik} method for computing the likelihood and thus is free to choose a suitable optimization method for his problem.

The function \code{KFS} computes the filtered (one-step-ahead prediction) and smoothed estimates for states, signals, and the values of the inverse link function (expected value $\mu$ or probability $\pi$) in a non-Gaussian case. For Gaussian models, disturbance smoothing is also available.

With \code{simulateSSM} the user can simulate the states, signals or disturbances of the Gaussian state space models given the model and the observations. If the model contains missing observations, these can also be simulated by \code{simulateSSM} in a similar way. It is also possible to simulate states from predictive distributions $p(\alpha_t|y_1,\ldots,y_{t-1})$, $t=1,\ldots,n$. For these simulations, instead of using marginal distributions $N(a_t,P_t)$, \pkg{KFAS} uses a modification of \citet{DK2002}, where smoothing is replaced by filtering.

For non-Gaussian models, \code{importanceSSM} returns the states or signals simulated from the approximating Gaussian model, and the corresponding weights $w_i$, which can then be used to compute arbitrary functions of the states or signals.

There are several \code{S3} methods available for \code{SSModel} and \code{KFS} objects. For both objects, simple \code{print} methods are provided, and for \code{SSModel} objects there is the \code{logLik} method. The \code{predict} method is for computing  the point predictions together with confidence or prediction intervals. The extraction operator \code{[} for extracting and replacing the subsets of model elements is available for the \code{SSModel} class. Using  this method when modifying the model is suggested instead of a common list extractor \code{$}, as the latter can accidentally modify the dimensions of the corresponding model matrices. A simple \code{plot} method for residual inspection is also provided.

For the \code{KFS} object, the methods \code{residuals}, \code{rstandard}, and \code{hatvalues} are provided. Also, a function \code{signal} can be used for extracting subsets of signals from \code{KFS} objects, for example, the part of $Z_t\alpha_t$ that corresponds to the regression part of the model.

Methods \code{coef} and \code{fitted} for the quick extraction of state or mean estimates are also available for \code{KFS} and \code{SSModel} objects.

\section{Constructing common state space models with KFAS}\label{models}

This section presents some typical models which can be formulated in a state space form. More examples can be found on the main help page of \pkg{KFAS} by typing \code{?KFAS} after the package is loaded via \code{"library("KFAS")}. These examples include most of the examples presented in \citet{DK2012}. Additional examples illustrating the functionality of \pkg{KFAS} can be found from the documentation of the particular functions. 

All the auxiliary functions used in the formula argument of the function \code{SSModel} have some common arguments which are not directly related to the system matrices of the corresponding component. In complex multivariate models, an important argument is \code{index}, which defines the series for which the corresponding component is constructed. For example, if we have four time series ($p=4$), we may want to use a certain regression component only for series 2 and 4. In this case we use the argument \code{index = c(2,4)} when calling the appropriate \code{SSMregression} function. By default the index is \code{1:p} so the component is constructed for all series.

Another argument used in several auxiliary functions is \code{type}, which can take two possible values. The value \code{"distinct"} defines the component separately for each series defined by \code{index} (with covariance structure defined by the argument \code{Q}), whereas the value \code{"common"} constructs a single component which applies to all series defined by \code{index}. For example, we can define distinct random walk components for all series together with a covariance matrix which captures the dependencies of the different series, or we can define just a single random walk component which is common to all series.

\subsection{Structural time series}

A structural time series refers to the class of state space models where the observed time series is decomposed into several underlying components, such as trend and seasonal effects. The basic structural time series model is of the form
\begin{equation}
\begin{aligned}\label{structural}
y_t       &= \mu_t + \gamma_t + c_t + \epsilon_t, \quad \epsilon_t \sim N(0, H_t),\\
\mu_{t+1} &= \mu_t + \nu_t + \xi_t, \quad \xi_t \sim N(0, Q_{\textrm{level},t}), \\
\nu_{t+1} &= \nu_t + \zeta_t, \quad \zeta_t \sim N(0, Q_{\textrm{slope},t}),
\end{aligned}
\end{equation}
where $\mu_t$ is the trend component, $\gamma_t$ is the seasonal component and $c_t$ is the cycle component. The seasonal component with period $s$ can be defined in a dummy variable form
\begin{equation*}
\gamma_{t+1} = -\sum_{j=1}^{s-1}\gamma_{t+1-j} + \omega_t, \quad \omega_t \sim N(0,Q_{\textrm{seasonal},t}),
\end{equation*}
or in a trigonometric form, where
\begin{equation*}
\begin{aligned}
\gamma_{t} &= \sum_{j=1}^{\lfloor s/2 \rfloor}\gamma_{j,t}, \\
\gamma_{j,t+1} &= \gamma_{j,t} \cos\lambda_j + \gamma^{\ast}_{j,t} \sin\lambda_j + \omega_{j,t}, \\
\gamma^{\ast}_{j,t+1} &= - \gamma_{j,t} \sin\lambda_j + \gamma^{\ast}_{j,t} \cos\lambda_j + \omega^{\ast}_{j,t}, \quad j=1,\ldots, \lfloor s/2 \rfloor,
\end{aligned}
\end{equation*}
with $\omega_{j,t}$ and $\omega^{\ast}_{j,t}$ being independently distributed variables with $N(0, Q_{\textrm{seasonal},t})$ distribution and $\lambda_j = 2\pi j/s$.

The cycle component with period $s$ is defined as
\begin{equation*}
\begin{aligned}
c_{t+1} &= c_t\cos\lambda_c + c_t^\ast\sin\lambda_c + \omega_t,\\
c^\ast_{t+1} &= - c_t\sin\lambda_c + c_t^\ast\cos\lambda_c + \omega^\ast_t,
\end{aligned}
\end{equation*}
with $\omega_t$ and $\omega^{\ast}_{t}$ being independent variables from $N(0, Q_{\textrm{cycle},t})$ distribution and frequency $\lambda_c = 2\pi/s$.

For non-Gaussian models the observation equation of~\eqref{structural} is replaced by $p(y_t|\theta_t)$, where $\theta_t = \mu_t + \gamma_t + c_t$. An additional Gaussian noise term $\epsilon_t$ can also be included in $\theta_t$ using the \code{SSMcustom} function (this is illustrated in Section~\ref{custom}). The general matrix formulation of structural time series can be found, for example, in \citet[Chapter 3]{DK2012}.

Three auxiliary functions, \code{SSMtrend}, \code{SSMcycle}, and \code{SSMseasonal}, for building structural time series are provided in \pkg{KFAS}. The argument \code{degree} of \code{SSMtrend} defines the degree of the polynomial component, where \code{1} corresponds to a local level model and \code{2} to a local linear trend model. Higher order polynomials can also be defined with larger values. Another important argument for \code{SSMtrend} is \code{Q}, which defines the covariance structure of the trend component. This is typically a list of $p \times p$ matrices (with $p$ being the number of series for which the component is defined), where the first matrix corresponds to the level component ($\mu$ in~\eqref{structural}), the second to the slope component $\nu$ and so forth.

The function \code{SSMcycle} differs from \code{SSMtrend} only by one argument. \code{SSMcycle} does not have argument \code{degree}, but instead it has argument \code{period} which defines the length of the cycle $c_t$. The same argument is also used in the function \code{SSMseasonal}, which contains also another important argument \code{sea.type}, which can be used to define whether the user wants a \code{dummy} or a \code{trigonometric} seasonal.

The example models of Sections~\ref{gaussian_example} and \ref{nongaussian_example} are special cases of the local linear trend model, where the variance of $\zeta_t$ is zero, and there are no seasonal or cycle components. Thus the Gaussian model of Section~\ref{gaussian_example} can be built with \pkg{KFAS} more easily by the following code:
\begin{Schunk}
\begin{Sinput}
R> model_structural <- SSModel(deaths / population ~ 
+      SSMtrend(degree = 2, Q = list(matrix(NA), matrix(0))), H = matrix(NA))
R> fit_structural <- fitSSM(model_structural, inits = c(0, 0), 
+    method = "BFGS")
R> fit_structural$model["Q"]
\end{Sinput}
\begin{Soutput}
, , 1

         [,1] [,2]
[1,] 4.256967    0
[2,] 0.000000    0
\end{Soutput}
\end{Schunk}

Here the state equation is defined using the \code{SSMtrend} auxiliary function without the need for an explicit definition of the corresponding system matrices. The intercept term is automatically omitted from the right side of the formula when the \code{SSMtrend} component is used, in order to keep the model identifiable. Here the unknown variance parameters are set to \code{NA}, so the default behaviour of the \code{fitSSM} function can be used for the parameter estimation.

\subsection{ARIMA models}

Another typical time series modelling framework are ARIMA models, which are also possible to define as a state space model. The auxiliary function \code{SSMarima} defines the ARIMA model using vectors \code{ar} and \code{ma}, which define the autoregressive and moving average coefficients, respectively. The function assumes that all series defined by the \code{index} have the same coefficients. The argument \code{d} defines the degree of differencing, and a logical argument \code{stationary} defines whether stationarity (after differencing) is assumed (if not, diffuse initial states are used instead of a stationary distribution). A univariate ARIMA($p$,$d$,$q$) model can be written as
\begin{equation*}
y^*_t = \phi_1 y^*_{t-1} + \ldots + \phi_p y^*_{t-p} + \xi_t + \theta_1 \xi_{t-1} + \ldots + \theta_q \xi_{t-q},
\end{equation*}
where $y^*_t = \Delta^d y_t$ and $\xi_t \sim \textrm{N}(0,\sigma^2)$. Let $r=\max(p,q+1)$. \pkg{KFAS} defines the state space representation of the ARIMA($p$,$d$,$q$) model with stationary initial distribution as
\begin{displaymath}
Z^\top =
\left( \begin{array}{c}
1_{d+1} \\
0 \\
\vdots \\
0
\end{array}\right),
H = 0,
T =
\left( \begin{array}{ccccc}
U_d & 1^\top_d & 0 & \cdots & 0\\
0 &\phi_1 & 1 & & 0 \\
\vdots &  &  & \ddots &  \\
\vdots & \phi_{r-1} & 0 &  & 1 \\
0 & \phi_{r} & 0 & \cdots & 0
\end{array} \right),
R =
\left( \begin{array}{c}
0_{d} \\
1 \\
\theta_1 \\
\vdots \\
\theta_{r-1}
\end{array}\right),
\end{displaymath}
\begin{displaymath}
\alpha_t =
\left( \begin{array}{c}
y_{t-1} \\
\vdots \\
\Delta^{d-1} y_{t-1}\\
y^*_t \\
\phi_2 y^*_{t-1} + \ldots + \phi_r y^*_{t-r+1} + \theta_1\eta_t+\ldots+\theta_{r-1}\eta_{t-r+2} \\
\vdots \\
\phi_r y^*_{t-1} + \theta_{r-1}\eta_t
\end{array}\right),
Q=\sigma^2,
\end{displaymath}
\begin{displaymath}
a_1 =
\left( \begin{array}{c}
0 \\
\vdots \\
0
\end{array}\right),
P_{*,1} =
\left( \begin{array}{cccccc}
0 & 0 \\
0 & S_r
\end{array} \right),
P_{\infty,1} =
\left( \begin{array}{cccccc}
I_{d} & 0 \\
0 & 0
\end{array} \right),
\eta_t=\xi_{t+1}
\end{displaymath}
where $\phi_{p+1}=\ldots=\phi_r=\theta_{q+1}=\ldots=\theta_{r-1}=0$, $1_{d+1}$ is a $1\times (d+1)$ vector of ones, $U_d$ is a $d \times d$ upper triangular matrix of ones and $S_r$ is the covariance matrix of stationary elements of $\alpha_1$. The elements of the initial state vector $\alpha_1$ which correspond to the differenced values $y_0, \ldots, \Delta^{d-1}y_0$ are treated as diffuse. The covariance matrix $S_r$ can be computed by solving the linear equation $(I-T \otimes T)\textrm{vec}(S_r) = \textrm{vec}(RR^\top)$ \citep[p.138]{DK2012}.

Note that the \code{arima} function from \pkg{stats} also uses the same state space approach for ARIMA modelling, although the handling of the intercept and possible covariates is done in a slightly different manner (see documentation of \code{arima} for details).

As an example, we again model the alcohol-related deaths but now use the ARIMA(0,1,1) model with drift: 
\begin{Schunk}
\begin{Sinput}
R> drift <- 1:length(deaths)
R> model_arima <- SSModel(deaths / population ~ drift + 
+      SSMarima(ma = 0, d = 1, Q = 1))
R> 
R> update_model <- function(pars, model) {
+    tmp <- SSMarima(ma = pars[1], d = 1, Q = pars[2])
+    model["R", states = "arima"] <- tmp$R
+    model["Q", states = "arima"] <- tmp$Q
+    model["P1", states = "arima"] <- tmp$P1
+    model
+  }
R> 
R> fit_arima <- fitSSM(model_arima, inits = c(0, 1), updatefn = update_model, 
+    method = "L-BFGS-B", lower = c(-1, 0), upper = c(1, 100))
R> fit_arima$optim.out$par
\end{Sinput}
\begin{Soutput}
[1] -0.4994891 16.9937888
\end{Soutput}
\end{Schunk}

In this case we need to supply the model updating function for \code{fitSSM} which updates our model definition based on the current values of the parameters we are estimating. Instead of manually altering the corresponding elements of the model, \code{update_model} uses \code{SSMarima} function for computation of relevant system matrices $R, Q$ and $P_1$. The estimated values for $\theta_1$ and $\sigma$ are \ensuremath{-0.5} and 17.

Comparing the results of our previous structural time series model and the estimated ARIMA model, we see that the estimated drift term and the log-likelihood are identical:
\begin{Schunk}
\begin{Sinput}
R> (out_arima <- KFS(fit_arima$model))
\end{Sinput}
\begin{Soutput}
Smoothed values of states and standard errors at time n = 39:
        Estimate  Std. Error
drift    0.8409    0.3446   
arima1  20.3008   13.1100   
arima2   1.3545    1.3898   
arima3   0.3031    0.6453   
\end{Soutput}
\begin{Sinput}
R> (out_structural <- KFS(fit_structural$model))
\end{Sinput}
\begin{Soutput}
Smoothed values of states and standard errors at time n = 39:
       Estimate  Std. Error
level  54.7532    2.1705   
slope   0.8409    0.3446   
\end{Soutput}
\begin{Sinput}
R> out_arima$logLik
\end{Sinput}
\begin{Soutput}
[1] -108.9734
\end{Soutput}
\begin{Sinput}
R> out_structural$logLik
\end{Sinput}
\begin{Soutput}
[1] -108.9734
\end{Soutput}
\end{Schunk}

This is not suprising given the well-known connections between structural time series and ARIMA models \citet{Harvey1989}.

\subsection{Linear and generalized linear models}

An ordinary linear regression model
\begin{equation*}
y_t = x^\top_t \beta + \epsilon_t, \quad t=1,\ldots,n,
\end{equation*}
where $\epsilon_t \sim N(0,\sigma^2)$, can be written as a Gaussian state space model by defining $Z_t=x^\top_t$, $H_t=\sigma^2$, $R_t=Q_t=0$ and $\alpha_t=\beta$. Assuming that the prior distribution of $\beta$ is defined as diffuse, the diffuse likelihood of this state space model corresponds to a restricted maximum likelihood (REML). Then the estimate for $\sigma^2$ obtained from \code{fitSSM} would be the familiar unbiased REML estimate of residual variance. It is important to notice that for this simple model numerical optimization is not needed, since we can estimate $\sigma^2$ by running the Kalman filter with $H_t=1$, which gives us
\begin{equation*}
\hat \sigma^2 = \frac{1}{\sum I(F_{\infty,t} = 0)} \sum_{t=1}^n I(F_{\infty,t} = 0) v_t^2/F_t,
\end{equation*}
which equals to the REML estimate of $\sigma^2$. The initial Kalman filter already provides correct estimates of $\beta$ as $a_{n+1}$, and running the Kalman filter again with $H_t=\sigma^2$ also gives the covariance matrix of $\hat \beta$ as $P_{n+1}$.

The extension from a linear model to a generalized linear model is straightforward as the basic theory behind the exponential family state space modelling can be formulated from the theory of generalized linear models (GLM) and can be thought of as extension to GLMs with additional dynamic structure. The iterative process of finding the approximating Gaussian model is equivalent with the famous iterative reweighted least squares (IRLS) algorithm \citep[p.~40]{MN1989}. If the model is ordinary GLM, the final estimates of regression coefficients $\beta$ and their standard errors coincide with maximum likelihood estimates obtained from ordinary GLM fitting. By adjusting the prior distribution for $\beta$ we can use \pkg{KFAS} also for the Bayesian analysis of Poisson and binomial regression (as those distributions do not depend on any additional parameters such as residual variance) with Gaussian prior.

A simple (generalized) linear model can be defined using \code{SSModel} without any auxiliary functions by defining the regression formula in the main part of the \code{formula}. For example, the following code defines a Poisson GLM which is identical to the one found on the help page of \code{glm}:
\begin{Schunk}
\begin{Sinput}
R> counts <- c(18, 17, 15, 20, 10, 20, 25, 13, 12)
R> outcome <- gl(3, 1, 9)
R> treatment <- gl(3, 3)
R> model_glm1 <- SSModel(counts ~ outcome + treatment, 
+    distribution = "poisson")
\end{Sinput}
\end{Schunk}
The previous model could also be defined using the auxiliary function \code{SSMregression}:
\begin{Schunk}
\begin{Sinput}
R> model_glm2 <- SSModel(counts ~ SSMregression(~ outcome + treatment), 
+    distribution = "poisson")
\end{Sinput}
\end{Schunk}

If our observations are multivariate, distinct regression components are defined for each of the series. For example, if counts \code{counts} above were a bivariate series, then both series would have ther own regression coefficients but the same covariate values. By using \code{SSMregression} explicitly, one could also define \code{type = "common"}, which would construct common regression coefficients for all series.

With \code{SSMregression} one can also define more complex regression models. The first argument of \code{SSMregression}, \code{rformula} can be used to provide a single formula or a list of formulas, where each component of the list contains the appropriate formula to be used for the corresponding series ($i$th formula in the list is used for the $i$th series defined by the argument \code{index}). When \code{rformula} is a list, the \code{data} argument of \code{SSMregression} can be a single data frame (or environment), or a list of such data objects. If \code{data} is a list, $i$th element of that list is used for $i$th formula, and if \code{data} is a single data frame or environment, the same data is used for all formulas. 

The state space approach makes it possible to extend classical GLMs in many ways. The extension to multivariate GLMs is straightforward, allowing, for example, the modelling of multiple groups of data where some of the model parameters are assumed to be identical between the groups, or where the number of explanatory variables differs between groups. For Gaussian models, a correlation of error terms $\epsilon$ between groups can also be incorporated. The use of dynamic GLM, where the regression coefficients follow a random walk process, can be defined by using argument \code{Q} in \code{SSMregression}. By manually altering the corresponding elements in the $T$ matrix, one can also define autoregressive behaviour for the coefficients. An additional parameter $u_t$ is defined separately for each observation, making it possible to define models where, for example, the dispersion parameter of a negative binomial model varies in time. One can also compute prediction intervals and other interesting measures efficiently via the importance sampling approach discussed in Section~\ref{exp}.

\subsection{Generalized linear mixed models}

Just like in GLM setting, it is also possible to write the generalized linear mixed model (GLMM) as a state space model. The difference between fixed and random effects lies in the initial state distribution; fixed effects are initialized via diffuse prior whereas random effects have proper variance defined by elements of $P_1$. Both types of states are automatically estimated by the Kalman filter, given the covariance structure of the random effects (and the residual variance or other parameters related to the distribution of the observation equation).

In practice, the mixed model formulation becomes quite cumbersome especially in hierarchical settings, but with large longitudinal settings it might still be useful to write a mixed model as a state space model, as it is then straightforward to add, for example, stochastic cycles or trends to the model. As an example I define a linear mixed model for the sleep deprivation study data from \pkg{lme4} \citep{lme4article,lme4} package as on the help page of the data. The data frame \code{sleepstudy} consists of three variables, the response variable \code{Reaction} (average reaction time), \code{Days} (number of days of sleep deprivation) and grouping variable \code{Subject}. First the response variable is restructured to a \code{matrix} (or \code{ts}) object:
\begin{Schunk}
\begin{Sinput}
R> library("lme4", quietly = TRUE)
R> y_split <- split(sleepstudy["Reaction"], sleepstudy["Subject"])
R> p <- length(y_split)
R> y <- matrix(unlist(y_split), ncol = p,
+    dimnames = list(NULL, paste("Subject", names(y_split))))
\end{Sinput}
\end{Schunk}

The data frame with explanatory variables is also split to a list where each list component corresponds to one group.
\begin{Schunk}
\begin{Sinput}
R> dataf <- split(sleepstudy, sleepstudy["Subject"])
\end{Sinput}
\end{Schunk}

The only explanatory variable \code{Days} in the data is identical to each Subject so the previous split of the data frame is not necessary, but illustrates the workflow for more complex data. 

We can now build the state space model by defining the common fixed part for each group (\code{SSMregression} function with argument \code{type = "common"}). Using the same function we can define the distinct random effect parts for each group, and the covariance structure of the random effects using the argument \code{P1} (the diffuse part \code{P1inf} is automatically set to zero for those states where the corresponding element in \code{P1} is nonzero). The function \code{.bdiag} from the \pkg{Matrix} package \citep{Matrix} is used for building a block diagonal covariance matrix for the random effects.
\begin{Schunk}
\begin{Sinput}
R> P1 <- as.matrix(.bdiag(replicate(p, matrix(NA, 2, 2), simplify = FALSE)))
R> model_lmm <- SSModel(y ~ -1 +
+      SSMregression(rep(list(~ Days), p), type = "common", data = dataf,
+        remove.intercept = FALSE) +
+      SSMregression(rep(list(~ Days), p), data = dataf,
+        remove.intercept = FALSE, P1 = P1),
+    H = diag(NA, p))
\end{Sinput}
\end{Schunk}
Note that in \code{SSMregression}, we use a lists of formulas (the first unnamed argument \code{rformula}) and data frames (argument \code{data}), where each component corresponds to one column of \code{y}. Thus we could use different formulas for different groups in more complex models. In this simple example the same model could be built with call
\begin{Schunk}
\begin{Sinput}
R> model_lmm2 <- SSModel(y ~ - 1 +
+      SSMregression(~ Days, type = "common", remove.intercept = FALSE) +
+      SSMregression(~ Days, remove.intercept = FALSE, P1 = P1),
+    H = diag(NA, p), data = data.frame(Days = 0:9))
\end{Sinput}
\end{Schunk}

Again we need to define the model updating function for \code{fitSSM}:
\begin{Schunk}
\begin{Sinput}
R> update_lmm <- function(pars, model) {
+    P1 <- diag(exp(pars[1:2]))
+    P1[1, 2] <- pars[3]
+    P1 <- crossprod(P1)
+    model["P1", states = 3:38] <- 
+      as.matrix(.bdiag(replicate(p, P1, simplify = FALSE)))
+    model["H"] <- diag(exp(pars[4]), p)
+    model
+  }
R> 
R> fit_lmm <- fitSSM(model_lmm, c(1, 1, 1, 5), update_lmm, method = "BFGS")
\end{Sinput}
\end{Schunk}

The estimated likelihood, variance/covariance parameters, and the estimates of fixed and random effects are practically identical to the ones obtained by \pkg{lmer} function of \pkg{lme4} package. The only major difference is in the estimation of conditional covariance matrices of the random effects, which is due to the fact that in \pkg{lme4} these matrices are computed conditionally on all the other model parameters, including the fixed effects \citep[p.28]{lme4article}. In \pkg{KFAS} the conditioning is only on the the numerically estimated variance/covariance parameters, and thus the resulting standard errors of random effects take account of the uncertainty of the estimation of fixed effects also.

In this example different groups were thought of as separate response vectors. In cases where the sample sizes in different groups are not equal, the same approach can be used after appropriately filling the data matrix with missing values. The corresponding \code{NA} values in covariates do not cause problems as they are not referenced in the Kalman filter. 

It is also possible to define the mixed model using univariate response and time-varying system matrices $T_t$ and $Q_t$. This reduces the state space and thus makes the model computationally more efficient, but adding other stochastic components to the model can be more problematic. For building such a model we need to use either the \code{customSSM} function for defining the corresponding time-varying system matrices, or we can use \code{SSMregression} as a starting point and alter the model components manually. This univariate approach is illustated on the main help page of \pkg{KFAS}.

\subsection{Arbitrary state space models}\label{custom}

By combining the auxialiary functions presented in the previous sections and possibly manually adjusting the resulting system matrices, a large amount of models can be constructed with relative ease. For cases where this is not sufficient or otherwise preferable, the auxiliary function \code{SSMcustom} can be used for the construction of arbitrary components by direct definition of the system matrices. As an example, we modify the Poisson model of Section~\ref{exp} by adding an additional white noise term which tries to capture possible overdispersion of the data. Our model for the Poisson intensity is now $u_t \exp(\mu_t + \epsilon_t)$ with
\begin{equation*}
\begin{aligned}
\mu_{t+1} &=  \mu_t + \nu + \eta_t,
\end{aligned}
\end{equation*}
where $\eta_{t} \sim N(0,\sigma^2_{\eta})$ as before, and $\epsilon_t ~ \sim N(0,\sigma^2_{\epsilon})$. This model can be written in a state space form by defining
\begin{Schunk}
\begin{Sinput}
R> model_poisson <- SSModel(deaths ~ SSMtrend(2, Q = list(NA, 0)) + 
+      SSMcustom(Z = 1, T = 0, Q = NA, P1 = NA), 
+    distribution = "poisson",  u = population)
\end{Sinput}
\end{Schunk}

As the model contains unknown parameters in \code{P1}, we need to provide a specific model updating function for \code{fitSSM}:
\begin{Schunk}
\begin{Sinput}
R> update_poisson <- function(pars, model) {
+    model["Q", etas = "level"] <- exp(pars[1])
+    model["Q", etas = "custom"] <- exp(pars[2])
+    model["P1", states = "custom"] <- exp(pars[2])
+    model
+  }
R> fit_poisson <- fitSSM(model_poisson, c(-3, -3), 
+    update_poisson, method = "BFGS")
R> fit_poisson$model["Q", etas = "level"]
\end{Sinput}
\begin{Soutput}
[1] 0.00316852
\end{Soutput}
\begin{Sinput}
R> fit_poisson$model["Q", etas = "custom"]
\end{Sinput}
\begin{Soutput}
[1] 0.002506342
\end{Soutput}
\end{Schunk}

From Figure~\ref{fig:customexample_output} we see that the Gaussian structural time series model and the Poisson structural time series model with additional white noise produce nearly indistinguishable estimates of the smoothed trend $\mu_t$. This is due to the relatively high intensity of the Poisson process.

\begin{Schunk}
\begin{figure}[!ht]

{\centering \includegraphics[width=0.7\linewidth]{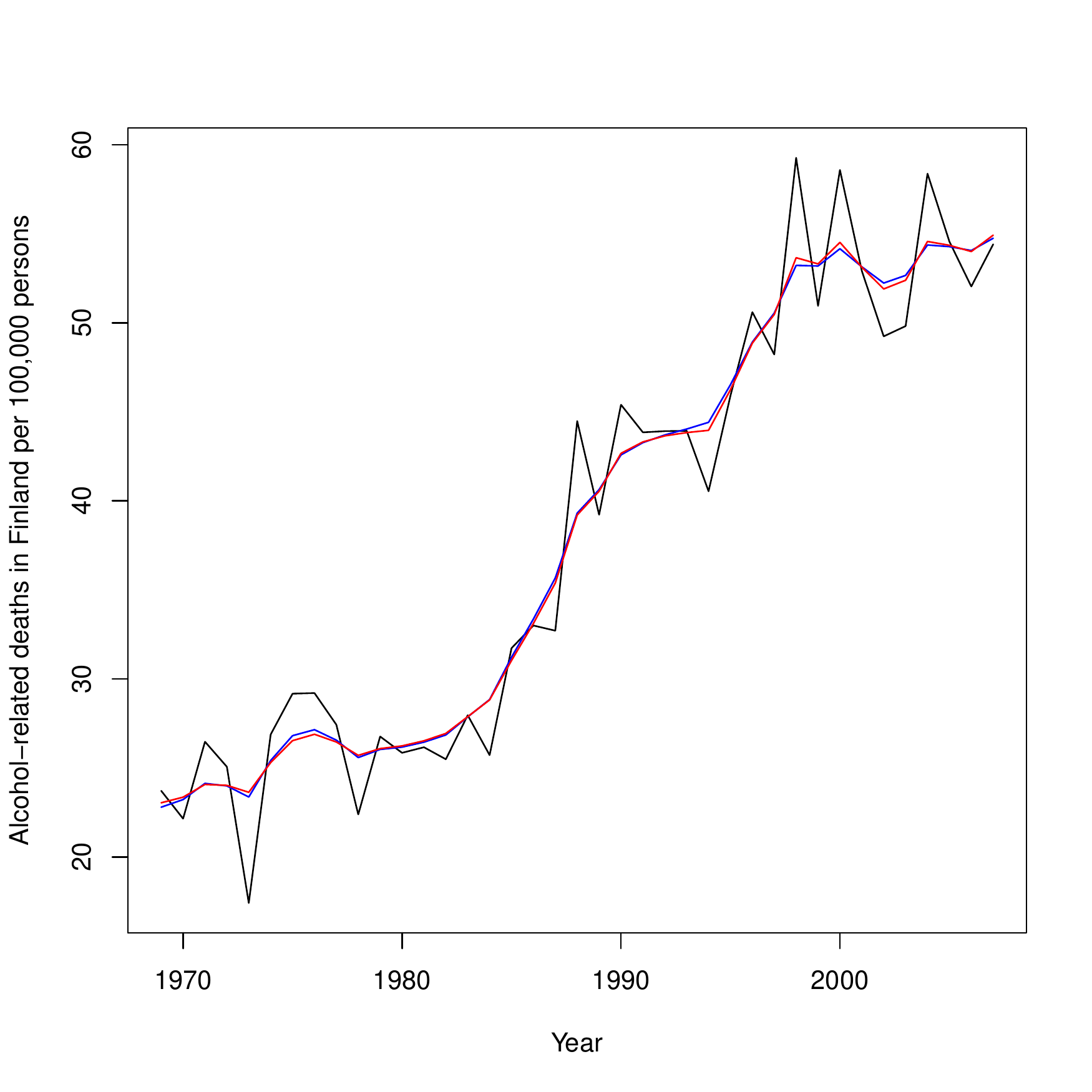} 

}

\caption[Alcohol-related deaths in Finland (black line) with smoothed estimates from Gaussian model (blue) and Poisson model with additional noise (red)]{Alcohol-related deaths in Finland (black line) with smoothed estimates from Gaussian model (blue) and Poisson model with additional noise (red).}\label{fig:customexample_output}
\end{figure}
\end{Schunk}

\section{Illustration}\label{illustration}

I now illustrate the use of \pkg{KFAS} with a more complete example case than the previous examples. Again the data consists of alcohol-related deaths in Finland, but now four age groups, 30--39, 40--49, 50--59 and 60--69, are modelled together as a multivariate Poisson model. The death counts and yearly population sizes in corresponding age groups are available for the years 1969--2012, but as an illustration, we only use the data  until 2007, and make predictions for the years 2008--2013. Figure~\ref{fig:alcoholPlot1} shows the number of deaths per 100,000 persons for all age groups.

\begin{Schunk}
\begin{Sinput}
R> data("alcohol")
R> colnames(alcohol)
\end{Sinput}
\begin{Soutput}
[1] "death at age 30-39"      "death at age 40-49"     
[3] "death at age 50-59"      "death at age 60-69"     
[5] "population by age 30-39" "population by age 40-49"
[7] "population by age 50-59" "population by age 60-69"
\end{Soutput}
\begin{Sinput}
R> ts.plot(window(alcohol[, 1:4] / alcohol[, 5:8], end = 2007), col = 1:4,
+    ylab = "Alcohol-related deaths in Finland per 100,000 persons", 
+    xlab = "Year")
R> legend("topleft",col = 1:4, lty = 1, legend = colnames(alcohol)[1:4])
\end{Sinput}
\begin{figure}[!ht]

{\centering \includegraphics[width=0.7\linewidth]{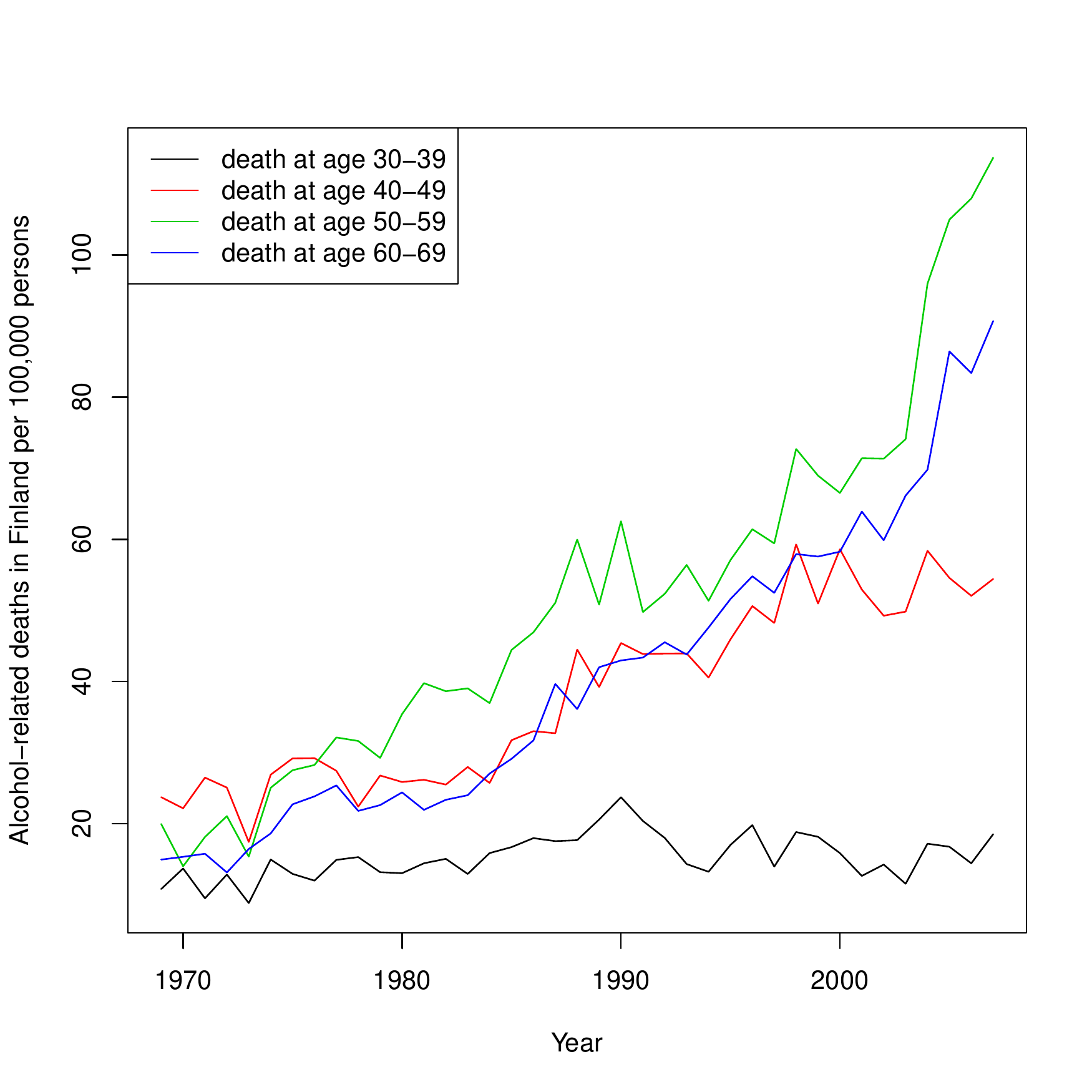} 

}

\caption[Alcohol-related deaths per 100,000 persons in Finland in 1969--2007 for four age groups]{Alcohol-related deaths per 100,000 persons in Finland in 1969--2007 for four age groups.}\label{fig:alcoholPlot1}
\end{figure}
\end{Schunk}
Here I choose a multivariate extension of the Poisson model used in Section~\ref{custom}:
\begin{equation}\label{alcohol}
\begin{aligned}
p(y_t|\theta_t) &= \textrm{Poisson}(u_t e^{\theta_t}), \quad u_t = \textrm{population}_t,\\
\theta_t &= \mu_t + \epsilon_t, \quad \epsilon_t \sim N(0, Q_{\textrm{noise}}),\\
\mu_{t+1} &= \mu_t + \nu_t + \xi_t, \quad \xi_t \sim N(0, Q_{\textrm{level}}),\\
\nu_{t+1} &= \nu_t.
\end{aligned}
\end{equation}

Here $\mu_t$ is the random walk with drift component, $\nu_t$ is a constant slope and $\epsilon_t$ is an additional white noise component which captures the extra variation of the series. I make no restrictions for the covariance structures of the level or the noise component.

The model~\eqref{alcohol} can be constructed with \pkg{KFAS} as follows.
\begin{Schunk}
\begin{Sinput}
R> alcoholPred <- window(alcohol, start = 1969, end = 2007)
R> model <- SSModel(alcoholPred[, 1:4] ~
+      SSMtrend(2, Q = list(matrix(NA, 4, 4), matrix(0, 4, 4))) +
+      SSMcustom(Z = diag(1, 4), T = diag(0, 4), Q = matrix(NA, 4, 4),
+        P1 = matrix(NA, 4, 4)), distribution = "poisson",
+    u = alcoholPred[, 5:8])
\end{Sinput}
\end{Schunk}

The updating function for \code{fitSSM} is
\begin{Schunk}
\begin{Sinput}
R> updatefn <- function(pars, model, ...){
+    Q <- diag(exp(pars[1:4]))
+    Q[upper.tri(Q)] <- pars[5:10]
+    model["Q", etas = "level"] <- crossprod(Q)
+    Q <- diag(exp(pars[11:14]))
+    Q[upper.tri(Q)] <- pars[15:20]
+    model["Q", etas = "custom"] <- model["P1", states = "custom"] <- 
+      crossprod(Q)
+    model
+  }
\end{Sinput}
\end{Schunk}

We can estimate the model parameters first without simulation, and then using those estimates as initial values run the estimation procedure again with importance sampling. In this case, the results obtained from the importance sampling step are practically identical with the ones obtained from the initial step.
\begin{Schunk}
\begin{Sinput}
R> init <- chol(cov(log(alcoholPred[, 1:4] / alcoholPred[, 5:8])) / 10)
R> fitinit <- fitSSM(model, updatefn = updatefn,
+    inits = rep(c(log(diag(init)), init[upper.tri(init)]), 2),
+    method = "BFGS")
R> -fitinit$optim.out$val
\end{Sinput}
\begin{Soutput}
[1] -704.8052
\end{Soutput}
\begin{Sinput}
R> fit <- fitSSM(model, updatefn = updatefn, inits = fitinit$optim.out$par,
+    method = "BFGS", nsim = 250)
R> -fit$optim.out$val
\end{Sinput}
\begin{Soutput}
[1] -704.8034
\end{Soutput}
\end{Schunk}

Using the model extraction method for the fitted models, we can check the estimated covariance and correlation matrices:
\begin{Schunk}
\begin{Sinput}
R> varcor <- fit$model["Q", etas = "level"]
R> varcor[upper.tri(varcor)] <- cov2cor(varcor)[upper.tri(varcor)]
R> print(varcor, digits = 2)
\end{Sinput}
\begin{Soutput}
       [,1]    [,2]   [,3]  [,4]
[1,] 0.0074 0.66022 0.8062 0.856
[2,] 0.0028 0.00239 0.1654 0.711
[3,] 0.0040 0.00047 0.0034 0.755
[4,] 0.0033 0.00156 0.0020 0.002
\end{Soutput}
\begin{Sinput}
R> varcor <- fit$model["Q", etas = "custom"]
R> varcor[upper.tri(varcor)] <- cov2cor(varcor)[upper.tri(varcor)]
R> print(varcor, digits = 2)
\end{Sinput}
\begin{Soutput}
        [,1]    [,2]    [,3]    [,4]
[1,] 0.00537 0.73118 0.75627 8.0e-01
[2,] 0.00315 0.00346 0.99924 9.9e-01
[3,] 0.00295 0.00313 0.00283 1.0e+00
[4,] 0.00043 0.00043 0.00039 5.4e-05
\end{Soutput}
\end{Schunk}

Parameter estimation of a state space model is often a difficult task, as the likelihood surface contains multiple maxima, thus making the optimization problem highly dependent on the initial values. Often the unknown parameters are related to the unobserved latent states, such as the covariance matrix in this example, with little a priori knowledge. Therefore, it is challenging to guess good initial values, especially in more complex settings. Thus, multiple initial value configurations possibly with several different type of optimization routines is recommended before one can be reasonably sure that proper optimum is found. Here we use the covariance matrix of the observed series as initial values for the covariance structures.

Another issue in the case of non-Gaussian models is the fact that the likelihood computation is based on iterative procedure which is stopped using some stopping criteria (such as the relative change of log-likelihood), so the log-likelihood function actually contains some noise. This in turn can affect the gradient computations in methods like BFGS and can in theory give unreliable results. Using derivative free method like Nelder-Mead is therefore sometimes recommended. On the other hand, BFGS is usually much faster than Nelder-Mead, and thus I prefer to try BFGS first at least in preliminary analysis.

Using the function \code{KFS} we can compute the smoothed estimates of states:

\begin{Schunk}
\begin{Sinput}
R> out <- KFS(fit$model, nsim = 1000)
R> out
\end{Sinput}
\begin{Soutput}
Smoothed values of states and standard errors at time n = 39:
                          Estimate    Std. Error
level.death at age 30-39   2.8559160   0.0784371
slope.death at age 30-39   0.0107142   0.0137135
level.death at age 40-49   4.0313117   0.0423763
slope.death at age 40-49   0.0237188   0.0076318
level.death at age 50-59   4.7578026   0.0398295
slope.death at age 50-59   0.0503715   0.0095850
level.death at age 60-69   4.4938371   0.0332897
slope.death at age 60-69   0.0482386   0.0072090
custom1                   -0.0004021   0.0603946
custom2                   -0.0195488   0.0408846
custom3                   -0.0169493   0.0370236
custom4                   -0.0021345   0.0051427
\end{Soutput}
\end{Schunk}

From the output of \code{KFS} we see that the slope term is not significant in the first age group. For time-varying states we can easily plot the estimated level and noise components, which shows clear trends in three age groups and highly correlated additional variation in all groups:

\begin{Schunk}
\begin{Sinput}
R> plot(coef(out, states = c("level", "custom")), main = "Smoothed states",
+    yax.flip = TRUE)
\end{Sinput}
\begin{figure}[!ht]

{\centering \includegraphics[width=\linewidth]{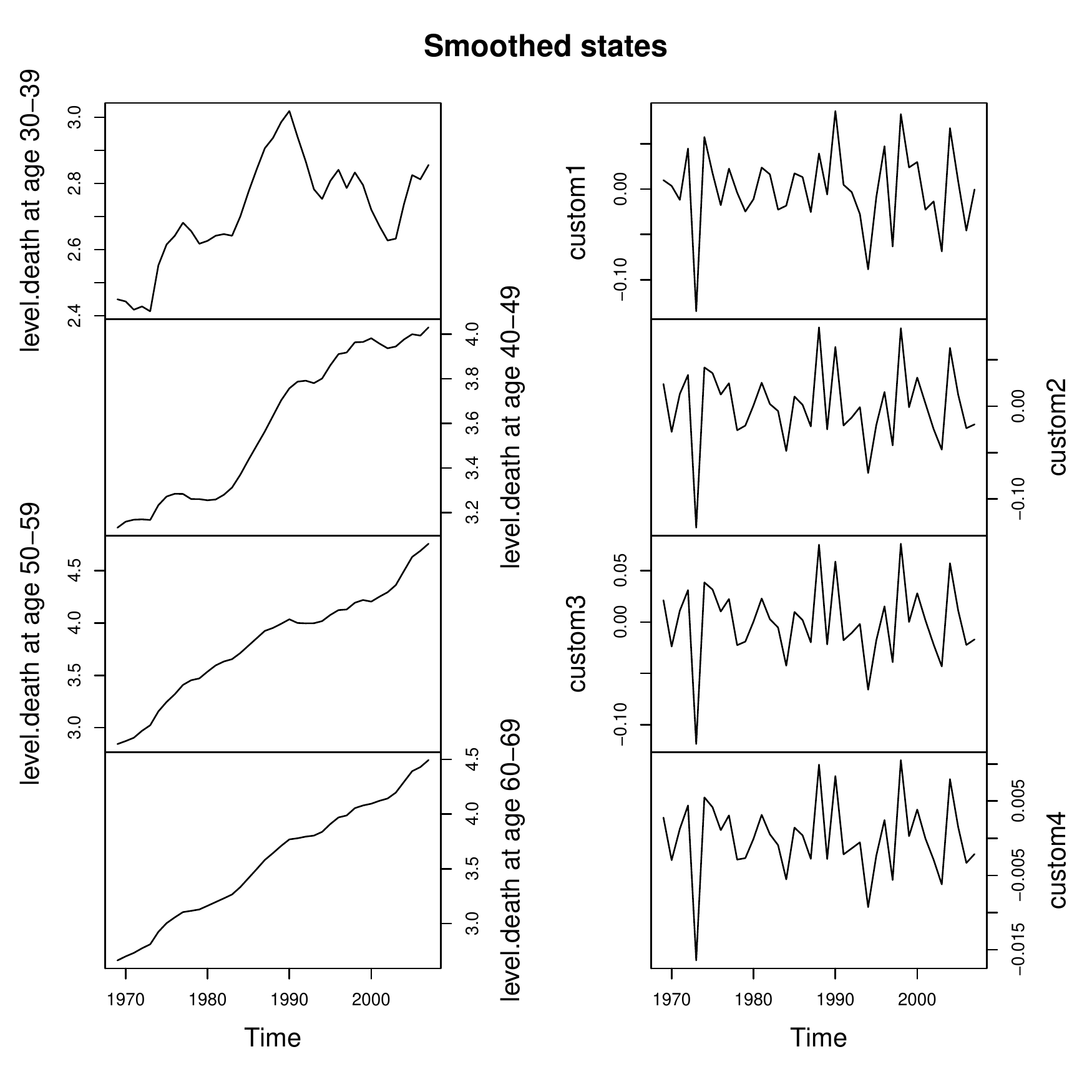} 

}

\caption[Smoothed level and white noise components]{Smoothed level and white noise components.}\label{fig:states}
\end{figure}
\end{Schunk}

Note the large drop in the noise component in Figure~\ref{fig:states}, which relates to a possible outlier in 1973 of the mortality series. As an illustration of model diagnostics, we compute recursive residuals for our model and check whether there is autocorrelation left in the residuals (Figure~\ref{fig:diagnostics1}).

\begin{Schunk}
\begin{Sinput}
R> res <- rstandard(KFS(fit$model, filtering = "mean", smoothing = "none",
+    nsim = 1000))
R> acf(res, na.action = na.pass)
\end{Sinput}
\begin{figure}[!ht]

{\centering \includegraphics[width=\linewidth]{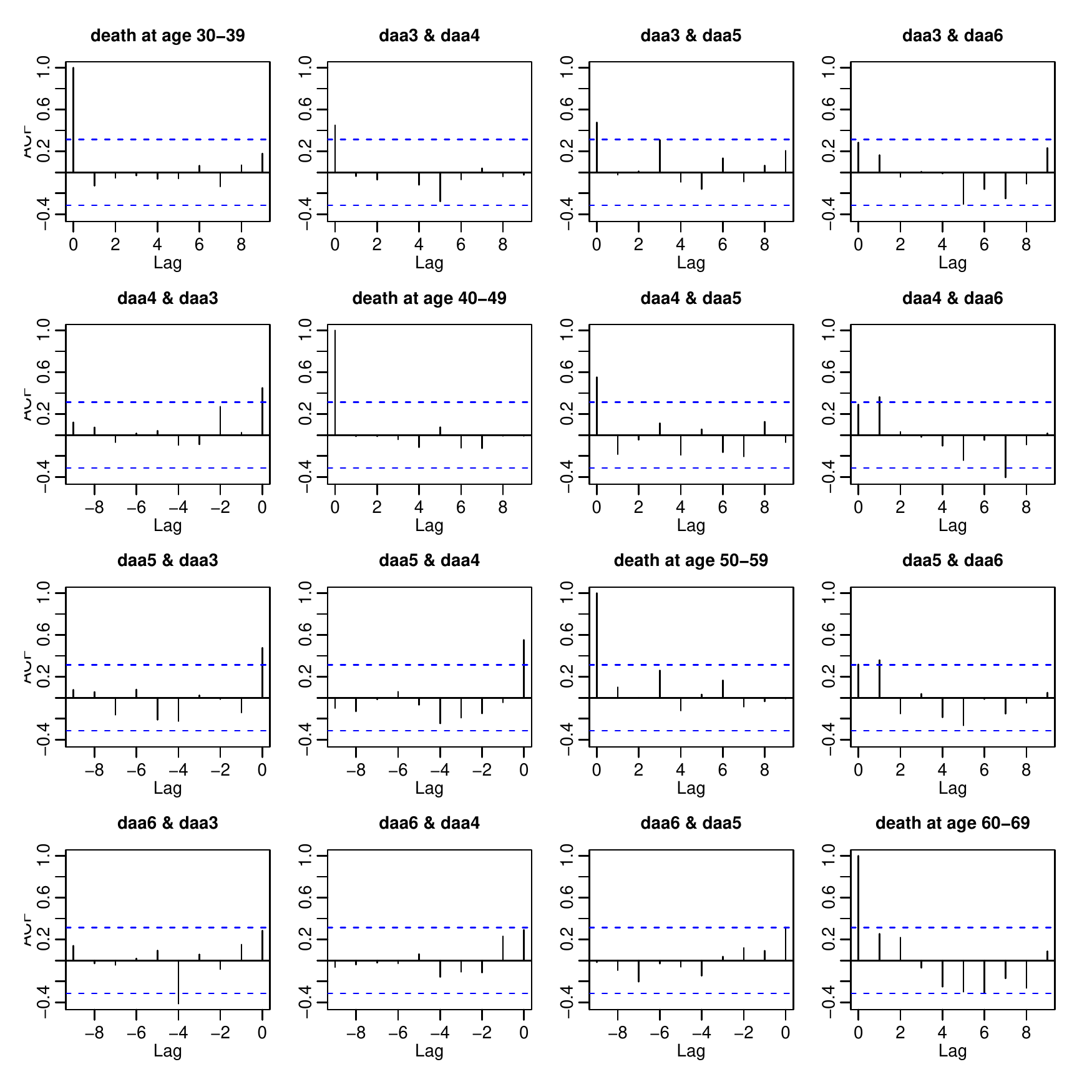} 

}

\caption[Autocorrelations and cross-correlations of recursive residuals]{Autocorrelations and cross-correlations of recursive residuals.}\label{fig:diagnostics1}
\end{figure}
\end{Schunk}

We see occasional lagged cross-correlation between the residuals, but overall we can be relatively satisfied with our model.

We can now predict the intensity $e^{\theta_t}$ of alcohol-related deaths per 100,000 persons for each age group for years 2008--2013 using our estimated model. As our model is time varying (\code{u} varies), we need to provide the model for the future observations via \code{newdata} argument. In this case we can use the \code{SSMcustom} function and provide all the necessary system matrices at once, together with constant \code{u = 1} (our signal $\theta$ is already scaled properly as the original $u_t$ was the population per 100,000 persons).

\begin{Schunk}
\begin{Sinput}
R> pred <- predict(fit$model,
+    newdata = SSModel(ts(matrix(NA, 6, 4), start = 2008) ~ -1 +
+        SSMcustom(Z = fit$model$Z, T = fit$model$T, R = fit$model$R,
+          Q = fit$model$Q), u = 1, distribution = "poisson"),
+    interval = "confidence", nsim = 10000)
\end{Sinput}
\end{Schunk}
\begin{Schunk}
\begin{Sinput}
R> trend <- exp(signal(out, "trend")$signal)
R> par(mfrow = c(2, 2), mar = c(2, 2, 2, 2) + 0.1, oma = c(2, 2, 0, 0))
R> for (i in 1:4) {
+    ts.plot(alcohol[, i]/alcohol[, 4 + i], trend[, i], pred[[i]],
+      col = c(1, 2, rep(3, 3)), xlab = NULL, ylab = NULL,
+      main = colnames(alcohol)[i])
+  }
R> mtext("Number of alcohol related deaths per 100,000 persons in Finland",
+    side = 2, outer = TRUE)
R> mtext("Year", side = 1, outer = TRUE)
\end{Sinput}
\begin{figure}[!ht]

{\centering \includegraphics[width=\linewidth]{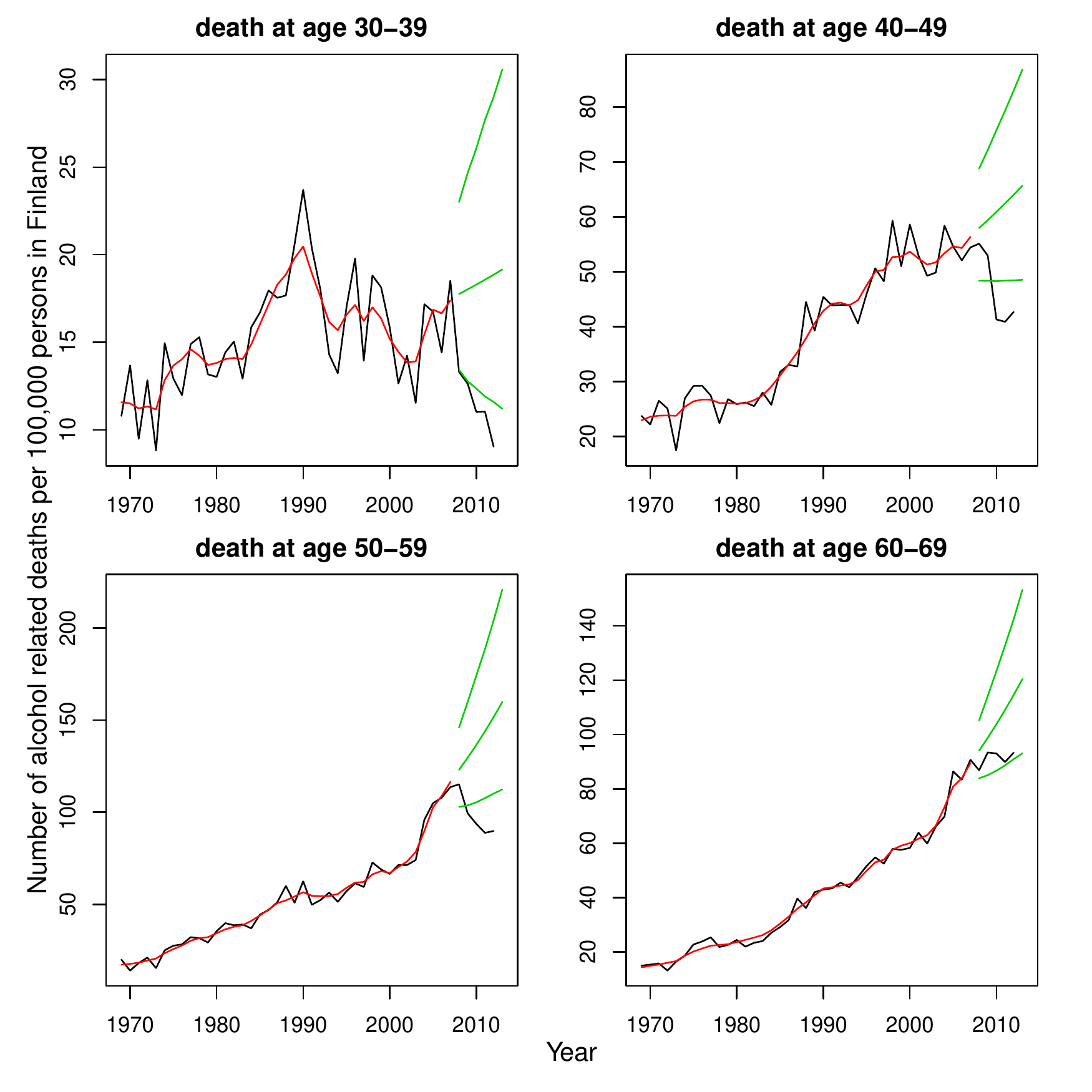} 

}

\caption[Observed number of alcohol related deaths per 100,000 persons in Finland (black), fitted values (red) and intensity predictions for years the 2008--2013 together with 95\% prediction intervals (green)]{Observed number of alcohol related deaths per 100,000 persons in Finland (black), fitted values (red) and intensity predictions for years the 2008--2013 together with 95\% prediction intervals (green).}\label{fig:predictplot}
\end{figure}
\end{Schunk}

Figure~\ref{fig:predictplot} shows the observed deaths, smoothed trends for 1969--2007, and intensity predictions for 2008--2013 together with 95\% prediction intervals for intensity. When we compare our predictions with true observations, we see that in reality the number of deaths slightly increased in the oldest age group (ages 60--69), whereas at another age they decreased substantially during the forecasting period. This is partly explained by the fact that during this period the total alcohol consumption decreased almost monotonically, which in turn might have been caused by the increase in taxation of alcohol in 2008, 2009 and 2012.

\clearpage 
\section{Other packages for non-Gaussian time series modelling}\label{otherpackages}

There are also other packages in CRAN which can be used for modelling non-Gaussian time series data. Package \pkg{pomp} \citep{pomp, pomparticle} offers functions for the inference of state space models with non-Gaussian and non-linear observation and state equations via particle filtering methods. The particle filtering approach makes \pkg{pomp} applicable to an even more broader class of models than \pkg{KFAS}, but the learning curve for using \pkg{pomp} is relatively high, as the user must write his or her own functions (preferably in \proglang{C}) for measurement and state process simulation as well as likelihood evaluation. Another package suitable for state space modelling is \pkg{INLA} \citep{inla, INLAarticle} (not available on CRAN), which can be used for Bayesian analysis via integrated nested Laplace approximation technique. Although it is often used in spatial modelling via Gaussian random fields, it can also be used for certain temporal state space models where the state transitions are Gaussian.

\pkg{KFAS} is based on a parameter-driven approach where the latent states $\alpha_t$ evolve in time as stochastic processes with the noise term $\eta_t$ which does not depend on the past observations or covariates. This approach offers a flexible and conceptually simple way of introducing multiple types of latent structures into the model. In contrast, in the observation-driven approach the state equation is defined using the past observations and possibly other covariates. This makes the states perfectly predictable (one-step-ahead) given the past information, allowing closed-form evaluation of the likelihood, and thus leading to computational gains compared to simulation-based estimation methods used in the parameter-driven approach for non-Gaussian models. Both approaches have their merits, see, for example, \citep{Koopman2015} for a comparison of parameter-driven and observation-driven approaches in a complex non-Gaussian non-linear setting.

\pkg{acp} \citep{acp} is a compact package based on the observation-driven approach for count data regression via Autoregressive Conditional Poisson (ACP) processes. In the ACP models the mean of the Poisson process is assumed to depend on the previous values of the observations and the previous values of the mean (which in turn can depend on covariates). A more general framework to the observation-driven approach for time series regression is implemented in the package \pkg{glarma} \citep{glarma}, which implements Generalized Linear Autoregressive Moving Average models (GLARMA) supporting Poisson, binomial and negative binomial distributions. Package \pkg{tscount} \citep{tscount} offers similar functionality using Poisson and negative binomial distributions with a closely related theoretic framework. All of these three packages assume univariate responses. 

The scope of the packages \pkg{gamlls} \citep{gamlss} and \pkg{VGAM} \citep{VGAM} is mainly on complex non-time series data, but they also have some capabilities for non-Gaussian time series modelling. Package \pkg{gamlss.util} \citep{gamlss.util} extends \pkg{gamlls} with the function \code{garmaFit} for univariate time series regression via GARMA models, which are closely related to the GLARMA models of the \pkg{glarma} package. A large number of distributions are supported. The GARMA models can also be estimated with \pkg{VGAM} \citep{VGAM} package, which contains the function \code{garma} for the estimation of GARMA models, but the documentation warns that the function is very unpolished.

Time series of counts often exhibit overdispersion or an excess amount of zeroes. Although previously mentioned packages can deal with these issues to some extent, there are also packages on CRAN designed specifically for these type of problems. Package \pkg{ZIM} \citep{ZIM} offers functions for both observation-driven and parameter-driven modelling of zero-inflated count series. For the parameter-driven approach a particle filtering approach is used. From the very scarce documentation of the package, it is not clear how the observation-driven approach is implemented. Package \pkg{tsintermittent} \citep{tsintermittent} contain forecasting methods for intermittent time series stemming, for example, from sales of slow moving items. Covariates are not supported.

Overall, there are multiple packages on CRAN which offer different approaches to non-Gaussian time series modelling, and preferring one package over another is likely dependent on the current problem in hand. For example, in some cases, time-dependency in the data can be more thought of as a nuisance which must be taken into account in order to make reliable inferences regarding the regression coefficients of the model. However, in some cases it can be that the interest is in the underlying latent time-varying processes itself. Due to the parameter-driven approach, packages such as \pkg{KFAS} and \pkg{pomp} can be used for the flexible modelling of, for example, a stochastic trend, seasonal and cyclic components. For packages such as \pkg{glarma} and \pkg{tscount} options are more limited because of the nature of the general model specification.

Time-varying regression coefficients and random effects can be incorporated to time series models with \pkg{INLA}, \pkg{KFAS} and \pkg{pomp}. These packages can also deal with missing observations in the response variable straightforwardly, whereas other packages do not seem to handle missing values properly. Most of the packages produce informative or non-informative error messages in the case of missing observations, whereas some just omit the missing time points of the data without taking account of the unevenness of the time points during the parameter estimation.

\subsection[Comparison to INLA]{Comparison to \pkg{INLA}}\label{comparison}

I will now briefly compare \pkg{KFAS} and \pkg{INLA}. As an illustration, we reanalyze the salmonella data analyzed by \citet{Margolin} which is available from \pkg{INLA}. The data consists of the number of revertant colonies of TA98 Salmonella with different doses of quinoline. We model the number of colonies as a Poisson GLMM with two explanatory variables and a random intercept term which tries to capture the overdispersion in the data. The codes for inference with \pkg{INLA} used here can be found from \url{http://www.r-inla.org/examples/volume-1/code-for-salm-example}. The \pkg{INLA} is not available at CRAN but can be downloaded from \url{http://www.math.ntnu.no/inla/R/stable}.

\begin{Schunk}
\begin{Sinput}
R> library("INLA", quietly = TRUE)
R> data("Salm")
R> mod.salm <- inla(y ~ log(dose + 10) + dose +
+      f(rand, model = "iid", param = c(0.001, 0.001)),
+    family = "poisson", data = Salm)
R> h.salm <- inla.hyperpar(mod.salm)
\end{Sinput}
\end{Schunk}

There are two ways to define the random intercept component in \pkg{KFAS}. The first one uses the \code{SSMregression} function and constructs a factor with 18 levels (one for each case) with non-diffuse initial variance $\sigma^2$. This gives 18 identically distributed time-invariant states, where each state corresponds to the random effect of one observation. Another option would be to use the \code{SSMcustom} function and define just one time-varying state as \code{SSMcustom(Z = 1, T = 0,  R = 1, Q = sigma2, a1 = 0, P1 = sigma2,  P1inf = 0)}. Both approaches give identical results. However, for large data the former approach is less efficient as the number of states depends on the number of observations. Nevertheless, we use the former approach here for illustration.

\begin{Schunk}
\begin{Sinput}
R> Salm$rand <- as.factor(Salm$rand)
R> model <- SSModel(y ~ log(dose + 10) + dose +
+      SSMregression(~ -1 + rand, P1 = diag(NA, 18), 
+        remove.intercept = FALSE),
+    data = Salm, distribution = "poisson")
R> 
R> updatefn <- function(pars,model,...){
+    diag(model["P1", states = 4:21]) <- exp(pars)
+    model
+  }
R> 
R> fit <- fitSSM(model, updatefn = updatefn, inits = -3, method = "BFGS",
+    nsim = 1000)
\end{Sinput}
\end{Schunk}

\begin{Schunk}
\begin{Sinput}
R> out <- KFS(fit$model, nsim = 10000)
R> out
\end{Sinput}
\begin{Soutput}
Smoothed values of states and standard errors at time n = 18:
                Estimate    Std. Error
(Intercept)      2.1674035   0.3519119
log(dose + 10)   0.3123230   0.0957658
dose            -0.0009783   0.0004249
rand1           -0.0998189   0.1929948
rand2            0.0811525   0.1905155
rand3            0.2962272   0.1852740
rand4           -0.1796443   0.1826173
rand5           -0.1210575   0.1804227
rand6           -0.0365042   0.1757863
rand7           -0.2997077   0.1738915
rand8           -0.0405697   0.1641720
rand9            0.1189599   0.1591311
rand10          -0.1649685   0.1664616
rand11           0.1202154   0.1569226
rand12           0.4328667   0.1490611
rand13          -0.1235595   0.1659587
rand14          -0.0274907   0.1650945
rand15           0.0274298   0.1624904
rand16          -0.2166174   0.1950416
rand17          -0.0434169   0.1915480
rand18           0.2725027   0.1914863
\end{Soutput}
\begin{Sinput}
R> h.salm$summary.fixed[, 1:2]
\end{Sinput}
\begin{Soutput}
                        mean           sd
(Intercept)     2.1671787697 0.3627732056
log(dose + 10)  0.3127852007 0.0987778011
dose           -0.0009794203 0.0004366782
\end{Soutput}
\begin{Sinput}
R> h.salm$summary.random$rand[, 2:3]
\end{Sinput}
\begin{Soutput}
          mean        sd
1  -0.09491396 0.1979799
2   0.08421357 0.1929301
3   0.29577465 0.1946343
4  -0.17165618 0.1860174
5  -0.11379450 0.1819493
6  -0.03076457 0.1776603
7  -0.29145696 0.1842334
8  -0.03433952 0.1653510
9   0.12229514 0.1607173
10 -0.15831766 0.1695020
11  0.12169606 0.1603516
12  0.42740226 0.1589524
13 -0.11885975 0.1685411
14 -0.02364225 0.1653818
15  0.03042788 0.1641417
16 -0.21238483 0.2025258
17 -0.04120592 0.1964929
18  0.26895855 0.1967189
\end{Soutput}
\begin{Sinput}
R> 1 / h.salm$summary.hyper[1]
\end{Sinput}
\begin{Soutput}
                         mean
Precision for rand 0.04825601
\end{Soutput}
\begin{Sinput}
R> fit$model["P1", states = 4]
\end{Sinput}
\begin{Soutput}
[1] 0.06554971
\end{Soutput}
\end{Schunk}

Although \pkg{INLA} uses a Bayesian approach, which takes account of the parameter estimation uncertainty, the results from \pkg{INLA} and \pkg{KFAS} are practically the same, even with such small data. The Kalman filtering with diffuse initialization still takes account of the uncertainty of the estimation of regression coefficients, so the differences here are related to the different prior definitions and the estimation of the hyperparameter $\sigma^2$, which is estimated as precision $1/\sigma^2$ by \pkg{INLA}. The estimate of $\sigma^2$ by \pkg{KFAS} is 0.066 whereas \pkg{INLA} gives $\sigma^2=0.048$. Note that changing the estimated $\sigma^2$ from \pkg{INLA} into the model estimated by \pkg{KFAS} produces a slightly lower log-likelihood value (-73.50 versus -73.66).

As \pkg{INLA} and \pkg{KFAS} are based on a different (although related) theoretical framework, the extensive study of their performances in terms of the computational efficiency and accuracy of results is somewhat pointless. Nevertheless, some remarks can be made. I feel that the biggest advantage of \pkg{INLA} is the Bayesian framework which allows us to take account of the parameter uncertainty in predictions and other inference. On the other hand, the computational burden related to the numerical integration over the hyperparameters can become infeasible as the number of hyperparameters increases. It is not uncommon to have a time series model with tens (or even hundreds) of parameters (such as multivariate structural time series or dynamic factor models). Of course, these same models can cause problems also to the maximum likelihood estimation, as noted in the Section~\ref{illustration}. Also the Bayesian approach eliminates the need for defining good initial values for the maximum likelihood estimation but the problem transforms into defining good priors for the same hyperparameters, which again is a non-trivial task in practice.

\section{Discussion}

State space models offer tools for solving a large class of statistical problems. Here I introduced an R package \pkg{KFAS} for linear state space modelling where the observations are from an exponential family. With such a general framework, different aspects of the modelling need to be taken into account. Therefore the focus of the package has been to provide reliable and relatively fast tools for multiple inference problems, such as maximum likelihood estimation, filtering, smoothing and simulation. Compared with the early versions of \pkg{KFAS}, constructing a state space model with simple components is now possible without an explicit definition of the system matrices by using the auxiliary functions and symbolic descriptions with the help of formula objects, which should greatly ease the use of the package.

Currently all the time consuming parts of \pkg{KFAS} are written in \proglang{Fortran}, which makes it relatively fast, given the general nature of the problems \pkg{KFAS} can handle. Still, converting the package to \proglang{C++} and \code{S4} classes with the help of \pkg{Rcpp} \citep{RcppA,RcppB} could result in potential improvements in terms of memory management, scalability and maintenance.

\section*{Acknowledgments}

The author wishes to thank Jukka Nyblom, Patricia Menendez, Spencer Graves, as well as the editor and two anonymous reviewers for the valuable comments and suggestions regarding the paper and the package. Comments, suggestions and bug reports from various users of \pkg{KFAS} over the years are also highly appreciated. The author has been financially supported by the Emil Aaltonen Foundation and the Academy of Finland research grant 284513.

\clearpage
\newpage
\appendix
\section{Appendix: Filtering and smoothing recursions}\label{appendix}

The following formulas summarize the Kalman filtering and smoothing formulas for diffuse and sequential case andare based on \citet{DK2012} and related articles. The original formulas are somewhat scattered between the references with slightly different notations. Therefore I have collected the equations used in \pkg{KFAS} to this Appendix.
\subsection{Filtering}
Denote
\begin{equation*}
\begin{aligned}
a_{t+1} &= \E(\alpha_{t+1}|y_t,\ldots,y_1) \quad \textrm{and}  \\
P_{t+1} &= \VAR(\alpha_{t+1}|y_t,\ldots,y_1).
\end{aligned}
\end{equation*}
The Kalman filter recursions for the general Gaussian model of form~\eqref{ssgeneral} are
\begin{equation*}
\begin{aligned}
v_t &= y_t - Z_t a_t \\
F_t &= Z_t P_t Z_t^\top + H_t \\
K_t &= P_t Z_t^\top \\
a_{t+1} &= T_t (a_t + K_t F^{-1}_t v_t) \\
P_{t+1} &= T_t (P_t -K_tF^{-1}_tK_t^\top)T_t^\top + R_t Q_t R_t,
\end{aligned}
\end{equation*}

For the univariate approach, the filtering equations are
\begin{equation*}
\begin{aligned}
v_{t,i} &= y_{t,i} - Z_{t,i} a_{t,i} \\
F_{t,i} &= Z_{t,i} P_{t,i} Z_{t,i}^\top + \sigma^2_{t,i} \\
K_{t,i} &= P_{t,i} Z_{t,i}^\top \\
a_{t,i+1} &= a_{t,i} + K_{t,i}F_{t,i}^{-1} v_{t,i} \\
P_{t,i+1} &= P_{t,i} -K_{t,i}K^\top_{t,i}F_{t,i}^{-1}\\
a_{t+1,1} &= T_t a_{t,p_t+1}\\
P_{t+1,1} &= T_t P_{t,p_t+1} T_t^\top + R_t Q_t R_t,
\end{aligned}
\end{equation*}
for $t=1,\ldots,n$ and  $i=1,\ldots,p_t$, where $v_{t,i}$ and $F_{t,i}$ are
scalars, $K_{t,i}$ is a column vector and $\sigma^2_{t,i}$ is the $i$th diagonal
element of $H_t$. It is possible that $F_{t,i}=0$, which case $a_{t,i+1} =
a_{t,i}$, $P_{t,i+1} = P_{t,i}$, and $v_{t,i}$ is computed as usual.

The diffuse filtering equations for univariate approach are
\begin{equation*}
\begin{aligned}
v_{t,i} &= y_{t,i} - Z_{t,i} a_{t,i} \\
F_{\ast,t,i} &= Z_{t,i} P_{\ast,t,i} Z_{t,i}^\top + \sigma^2_{t,i} \\
F_{\infty,t,i} &= Z_{t,i} P_{\infty,t,i} Z_{t,i}^\top \\
K_{\ast,t,i} &= P_{\ast,t,i} Z_{t,i}^\top \\
K_{\infty,t,i} &= P_{\infty,t,i} Z_{t,i}^\top,
\end{aligned}
\end{equation*}
and
\begin{equation*}
\begin{aligned}
a_{t,i+1} &= a_{t,i} + K_{\infty,t,i} v_{t,i}F_{\infty,t,i}^{-1} \\
P_{\ast,t,i+1} &= P_{\ast,t,i}
+K_{\infty,t,i}K^\top_{\infty,t,i}F_{\ast,t,i}F_{\infty,t,i}^{-2}
-(K_{\ast,t,i}K^\top_{\infty,t,i}+K_{\ast,t,i}K^\top_{\infty,t,i})F_{\infty,t,i}^{-1}\\
P_{\infty,t,i+1} &= P_{\infty,t,i}
-K_{\infty,t,i}K^\top_{\infty,t,i}F_{\infty,t,i}^{-1}\\
\end{aligned}
\end{equation*}
if $F_{\infty,t,i}>0$, and
\begin{equation*}
\begin{aligned}
a_{t,i+1} &= a_{t,i} + K_{\ast,t,i} v_{t,i}F_{\ast,t,i}^{-1} \\
P_{\ast,t,i+1} &= P_{\ast,t,i} -K_{\ast,t,i}K^\top_{\ast,t,i}F_{\ast,t,i}^{-1}\\
P_{\infty,t,i+1} &= P_{\infty,t,i},\\
\end{aligned}
\end{equation*}
if $F_{\infty,t,i}=0$. The transition equations from $t$ to $t+1$ are
\begin{equation*}
\begin{aligned}
a_{t+1,1} &= T_t a_{t,p_t+1}\\
P_{\ast,t+1,1} &= T_t P_{\ast,t,p_t+1} T_t^\top + R_t Q_t R_t\\
P_{\infty,t+1,1} &= T_t P_{\infty,t,p_t+1} T_t^\top.
\end{aligned}
\end{equation*}

\subsection{Smoothing}

Denote
\begin{equation*}
\begin{aligned}
\hat\alpha_{t} &= \E(\alpha_{t}|y_n,\ldots,y_1) \quad \textrm{and}  \\
V_{t} &= \VAR(\alpha_{t}|y_n,\ldots,y_1).
\end{aligned}
\end{equation*}

The smoothing algorithms of \pkg{KFAS} are based on the following recursions:
\begin{equation*}
\begin{aligned}
r_{t,i-1} &= Z_{t,i}^\top v_{t,i}F_{t,i}^{-1} + L_{t,i}^\top r_{t,i}, \\
r_{t-1,p_t} &= T_{t-1}^\top r_{t,0},\\
N_{t,i-1} &= Z_{t,i}^\top Z_{t,i}F_{t,i}^{-1} +L_{t,i}^\top N_{t,i}L_{t,i},\\
N_{t-1,p_t} &= T_{t-1}^\top N_{t,0}T_{t-1},\\
L_{t,i} &= I - K_{t,i}Z_{t,i}^\top F_{t,i}^{-1},
\end{aligned}
\end{equation*}
for $t=n,\ldots,1$ and $i=p_t,\ldots,1$, with $r_{n,p_n}=0$ and $N_{n,p_n}=0$.
From these recursions, we get state smoothing recursions
\begin{equation*}
\begin{aligned}
\hat\alpha_{t} &= a_{t,1} + P_{t,1}r_{t,0}\\
V_{t} &= P_{t,1}-P_{t,1}N_{t,0}P_{t,1},
\end{aligned}
\end{equation*}
and disturbance smoothing recursions
\begin{equation*}
\begin{aligned}
\hat \epsilon_{t,i} &= \sigma^2_{t,i}F_{t,i}^{-1}(v_{t,i} -K_{t,i}^\top r_{t,i}),\\
\VAR(\hat \epsilon_{t,i}) &= \sigma^2_{t,i} -
\sigma^4_{t,i}(F_{t,i}^{-1} -K_{t,i}^\top N_{t,i}K_{t,i}F_{t,i}^{-2}),\\
\hat \eta_{t} &= Q_tR_t^\top r_{t,0},\\
\VAR(\hat \eta_{t,i}) &= Q_tR^\top_t N_{t,0}R_tQ_t.
\end{aligned}
\end{equation*}

The recursions for diffuse phase are as follows.

\begin{equation*}
\begin{aligned}
L_{\infty,t,i} &= I - K_{\infty,t,i}Z_{t,i}F_{\infty,t,i}^{-1},\\
L_{t,i} &=
(K_{\infty,t,i}F_{t,i}F_{\infty,t,i}^{-1}-K_{t,i})Z_{t,i}F_{\infty,t,i}^{-1},\\
r_{0,t,i-1} &= L_{\infty,t,i}^\top r_{0,t,i}, \\
r_{1,t,i-1} &= Z_{t,i}^\top v_{t,i}F_{\infty,t,i}^{-1} + L_{\infty,t,i}^\top r_{1,t,i} +
L^\top_{t,i}r_{0,t,i}, \\
N_{0,t,i-1} &= L_{\infty,t,i}^\top N_{0,t,i}L_{\infty,t,i}\\
N_{1,t,i-1} &= L_{t,i}^\top N_{0,t,i}L_{\infty,t,i}+
L_{\infty,t,i}^\top N_{1,t,i}L_{\infty,t,i}+Z_{t,i}^\top Z_{t,i}F_{\infty,t,i}^{-1},\\
N_{2,t,i-1} &= L_{t,i}^\top N_{0,t,i}L_{t,i} + L_{\infty,t,i}^\top N_{1,t,i}L_{t,i} +
(L_{\infty,t,i}^\top N_{1,t,i}L_{t,i})^\top +
L_{\infty,t,i}N_{2,t,i}^\top L_{\infty,t,i} \\
&- Z_{t,i}^\top Z_{t,i}F_{t,i}F_{\infty,t,i}^{-2},\\
N_{t-1,p_t} &= T_{t-1}^\top N_{t,0}T_{t-1},
\end{aligned}
\end{equation*}
if $F_{\infty,t,i}>0$, and
\begin{equation*}
\begin{aligned}
L_{t,i} &= I - K_{t,i}Z_{t,i}F_{t,i}^{-1},\\
r_{0,t,i-1} &= Z_{t,i}^\top v_{t,i}F_{t,i}^{-1} + L_{t,i}^\top r_{0,t,i}, \\
r_{1,t,i-1} &= L_{t,i}^\top r_{1,t,i}, \\
N_{0,t,i-1} &= L_{t,i}^\top N_{0,t,i}L_{t,i}+Z_{t,i}^\top Z_{t,i}F_{t,i}^{-1}\\
N_{1,t,i-1} &= N_{1,t,i}L_{t,i}\\
N_{2,t,i-1} &= N_{2,t,i}L_{t,i},
\end{aligned}
\end{equation*}
otherwise. The transition from time $t$ to $t-1$ is by $N_{j,t-1,p_t} =
T_{t-1}^\top N_{j,t,0}T_{t-1}$ for $j=0,1,2$, and $r_{j,t-1,p_t} = T_{t-1}^\top r_{j,t,0}$
for $j=0,1$, with $r_{0,d,j}=r_{d,j}$, $r_{1,d,j}=0$, $N_{0,d,j}=N_{d,j}$, and
$N_{1,d,j}=N_{2,d,j}=0$, where $(d,j)$ is the last point of diffuse phase.
From these basic recursions, we get state smoothing recursions for diffuse phase
as
\begin{equation*}
\begin{aligned}
\hat\alpha_{t} &= a_{t,1} + P_{t,1}r_{0,t,0} + P_{\infty,t,1}r_{1,t,0},\\
V_{t} &= P_{t,1}-P_{t,1}N_{0,t,0}P_{t,1} -(P_{\infty,t,1}N_{1,t,0}P_{t,1})^\top
-P_{\infty,t,1}N_{1,t,0}P_{t,1} - P_{\infty,t,1}N_{2,t,0}P_{\infty,t,1},
\end{aligned}
\end{equation*}
and disturbance smoothing recursions
\begin{equation*}
\begin{aligned}
\hat \epsilon_{t,i} &= -\sigma^2_{t,i}K_{\infty,t,i}^\top r_{0,t,i},\\
\VAR(\hat \epsilon_{t,i}) &= \sigma^2_{t,i} -
\sigma^4_{t,i}K_{\infty,t,i}^\top N_{0,t,i}K_{\infty,t,i}F_{\infty,t,i}^{-2},
\end{aligned}
\end{equation*}
if $F_{\infty,t,i}>0$, and
\begin{equation*}
\begin{aligned}
\hat \epsilon_{t,i} &=
-\sigma^2_{t,i}(v_{t,i}F_{\infty,t,i}^{-1}-K_{t,i}^\top r_{0,t,i}),\\
\VAR(\hat \epsilon_{t,i}) &= \sigma^2_{t,i} -
\sigma^4_{t,i}(F_{t,i}^{-1}-K_{t,i}^\top N_{0,t,i}K_{t,i}F_{t,i}^{-2}),
\end{aligned}
\end{equation*}
if $F_{\infty,t,i}=0$. For $\hat\eta$, recursions are
\begin{equation*}
\begin{aligned}
\hat \eta_{t} &= Q_tR_t^\top r_{0,t,0},\\
\VAR(\hat \eta_{t,i}) &= Q_tR_t^\top N_{0,t,0}R_tQ_t.
\end{aligned}
\end{equation*}

\clearpage
\bibliography{jss2537}
\end{document}